\documentclass[twocolumn,showpacs,aps,epsfig]{revtex4}

\usepackage{graphicx}
\usepackage{epstopdf}
\usepackage{latexsym}

\usepackage[center]{subfigure}

\begin{document}

 \newcommand{\bq}{\begin{equation}}
 \newcommand{\eq}{\end{equation}}
 \newcommand{\bqn}{\begin{eqnarray}}
 \newcommand{\eqn}{\end{eqnarray}}
 \newcommand{\nb}{\nonumber}
 \newcommand{\lb}{\label}
\newcommand{\PRL}{Phys. Rev. Lett.}
\newcommand{\PL}{Phys. Lett.}
\newcommand{\PR}{Phys. Rev.}
\newcommand{\CQG}{Class. Quantum Grav.}

\title{Thermodynamics and classification of cosmological models in the Horava-Lifshitz theory of gravity}
 
\author{Anzhong Wang }
\email{anzhong_wang@baylor.edu}

\author{Yumei Wu }
\email{yumei_wu@baylor.edu}

\affiliation{GCAP-CASPER, Physics Department, Baylor University, Waco, 
TX 76798-7316}

\date{\today}

\begin{abstract}

We study thermodynamics of cosmological models in the Horava-Lifshitz theory of gravity, 
and systematically investigate the evolution of the   universe filled with a perfect fluid that 
has the equation of state $p=w\rho$, where $p$ and $\rho$ denote, respectively, the 
pressure and energy density of the fluid, and $w$ is an arbitrary real constant.  Depending 
on specific values of the free parameters involved in the models, we classify all of them into 
various cases. In each case the main properties of the evolution  are studied in detail, 
including the periods of deceleration and/or acceleration, and the existence of big bang, big 
crunch, and big rip singularities.  We pay particular attention on models that may give rise 
to a bouncing universe. 

\end{abstract}


\maketitle

\section{Introduction}
\renewcommand{\theequation}{1.\arabic{equation}} \setcounter{equation}{0}

Recently, Horava proposed a very attractive  quantum gravity
theory \cite{Horava}, motivated by the Lifshitz theory in solid state
physics \cite{Lifshitz}, for which the theory is usually referred
to as the Horava-Lifshitz (HL) theory. In the IR limit, the standard general 
relativity is recovered. By construction, it is non-relativistic and UV-renormalizable 
at least around the flat space. The effective speed of light diverges in the 
UV, and may potentially resolve the horizon problem without invoking 
inflationary scenario. Since the theory is brand new, detailed studies are highly 
demanded, before any definitive conclusions  are reached, although a great
 deal of efforts have already been devoted to these subjects, including the studies of
cosmology \cite{Cosmos,MNTY}, and black hole physics \cite{BHs}, among
others \cite{others}. In particular, in \cite{LMP,KK} the general field equations 
were derived. When applying them to cosmology, the complete set of field 
equations were given explicitly, from which it can be seen that the spatial 
curvature is enhanced by higher-order curvature terms, and this may allow 
us to address the flatness problem, and provide a bouncing cosmology 
\cite{Calcagni,brand}. It was also shown that almost-invariant super-horizon curvature 
perturbations can be produced \cite{Muk}. 

However, despite  these attractive features, the theory has already been
facing some challenging questions \cite{CNPS,LP}. In particular, it was shown
that the HL theory may suffer strong coupling problems due to the breaking of 
diffeomorphism invariance  \cite{CNPS}. As pointed out in \cite{Mukb},  these 
problems might be solved by preserving the prejactablity condition, as was 
done originally by Horava \cite{Horava}.  On the other hand, there are a
couple of reasons to abandon  the detailed balance condition. One is due to the 
fact that matter is not UV stable with this condition \cite{Calcagni}.
It also requires a non-zero (negative) cosmological constant in order to
have a correct coupling, and breaks parity in the purely gravitational sector \cite{SVW}. 
As shown explicitly in \cite{SVW}, the general theory can be properly formulated
without    the ``detailed balance" condition, but still keeping the projectability 
condition and preserving parity.  

In this paper, we  shall focus 
on  the thermodynamics of cosmological models in the HL theory of gravity
without detailed balance, and systematically investigate cosmological 
 models  for a perfect fluid with the equation 
of state $p = w \rho$, where $\rho$ and $p$ denote, respectively, the 
energy density and pressure of the fluid, and $w$ is an arbitrary real 
constant. We shall classify all these models according to the values 
of the parameters involved in the models, and study the evolution of the 
universe for each model. By doing so, we shall study the spacetime 
singularities, such as the big bang, big crunch and big rip, and 
identify the period(s) when the universe is accelerating or decelerating.
We pay particular attention on models that give rise to a bouncing 
universe. Specifically,  the paper is organized as follows: In Sec. II, we 
give a brief introduction to the HL theory, while in Sec. III, we 
present   the Friedmann-like field equations.  In Sec. IV, we study 
thermodynamics of  cosmological models, and
in Sec. V. we investigate the evolution of the universe when filled with a
perfect fluid with the equation of state $p = w\rho$. We study  these 
solutions case by case, and 
deduce the main properties of each model of the universe. Finally, 
in Sec. VI, we present our main conclusions.

It should be noted that  classification of a (non-relativistic)
matter coupled with a dark energy was considered recently in \cite{CTS05},
 in the framework of Einstein's theory, and the corresponding Penrose diagrams 
 were presented. Similar considerations were also carried
 out in a series of papers, and particular
attention was paid to obtain an effective potential $V(a)$  by fitting observational 
data sets \cite{szy}.  In \cite{Ha09}, such studies were generalized to a 
perfect fluid with the equation of state $p = w\rho$. In this paper, 
we shall generalize these studies to  the HL cosmology. 

\section{The Horava-Lifshitz Gravity Theory}

\renewcommand{\theequation}{2.\arabic{equation}} \setcounter{equation}{0}

In this section, we shall give a very brief introduction to the HL Theory. For 
detail, we refer readers to \cite{Horava,LMP,KK}. The dynamical variables
are $N, \; N_{i}$ and $g_{ij}\; (i, \; j = 1, 2, 3)$, in terms of which the 
metric takes the ADM form,
\bq
\lb{2.1}
ds^{2} = - N^{2}dt^{2} + g_{ij}\left(dx^{i} + N^{i}dt\right) 
     \left(dx^{j} + N^{j}dt\right),
\eq
where $N^{i} \equiv g^{ij}N_{j}$, and the coordinates $\left(t, x^{i}\right)$
scale as,
\bq
\lb{2.2}
t \rightarrow  {\ell}^{3} t,\;\;\;
x^{i}  \rightarrow {\ell} x^{i}.
\eq
Under the above scaling, the  dynamical variables scale as
\bq
\lb{2.3}
N \rightarrow  N,\;
g_{ij}   \rightarrow g_{ij},\;
N_{i} \rightarrow  {\ell}^{2} N_{i},\;
N^{i} \rightarrow  {\ell}^{-2} N^{i}. 
\eq

The total action of the HL theory consists of three parts, the kinetic part, $S_{k}$, 
the potential part, $S_{v}$, and the matter part, $S_{m}$, given by
\bqn
\lb{2.4}
S_{total} &=& S_{k} + S_{v} + S_{m}\nb\\
&=& \int{dt dx^{i} N \sqrt{g}\left({\cal{L}}_{k} + {\cal{L}}_{v}  
+ {\cal{L}}_{m}\right)},
\eqn
where $g$ is the determinant of the three-metric $g_{ij}$,
${\cal{L}}_{m} = {\cal{L}}_{m}\left(N,\; N_{i},\; g_{ij},\; \Phi\right)$
 the Lagrangian density of matter fields, denoted collectively by $\Phi$, and
\bqn
\lb{2.5}
{\cal{L}}_{k} &=& \alpha\left(K_{ij}K^{ij} - \lambda K^{2}\right),\nb\\ 
{\cal{L}}_{v} &=& \beta C_{ij}C^{ij} + \gamma\frac{\epsilon^{ijk}}{\sqrt{g}}
R_{il}\nabla_{j}R^{l}_{k} + \zeta R_{ij}R^{ij}\nb\\
& & + \eta R^{2} + \xi R + \sigma,
\eqn
where $\epsilon^{ijk}$ is the   antisymmetric tensor with
$\epsilon^{123} = 1$,  $\nabla_{k}$ denotes the covariant derivative with respect to
$g_{ij}$. $R_{ij}$  is the Ricci tensor of the three-metric $g_{ij}$, 
$R = g^{ij}R_{ij}$, and $C_{ij}$ and $K_{ij}$ are, respectively, the Cotton tensor
and extrinsic curvature, defined  by
\bqn
\lb{2.6}
C_{ij} &\equiv& \frac{\epsilon^{ijk}}{\sqrt{g}}\nabla_{k}\left(R^{j}_{l}
- \frac{1}{4} \delta^{j}_{l} R\right),\nb\\
K_{ij} &\equiv& \frac{1}{2N}\left(\dot{g}_{ij} - \nabla_{i}N_{j}
- \nabla_{j}N_{i}\right),
\eqn
where $\dot{g}_{ij} \equiv dg_{ij}/dt$. The constants $\alpha, \; \lambda,\;
\beta, \gamma,\; \zeta,\; \eta,\; \xi$ and $\sigma$ are coupling constants. Under the  
 ``detailed-balance" conditions, they are not independent, and are given by
\bqn
\lb{2.7}
\alpha &=& \frac{2}{\kappa^{2}},\;
\beta = -  \frac{\kappa^{2}}{2\omega^{4}},\;
\gamma =  \frac{\kappa^{2}\mu}{2\omega^{2}},\;
\zeta = -  \frac{\kappa^{2}\mu^{2}}{8},\nb\\
\eta &=& \frac{\kappa^{2}\mu^{2}(1-4\lambda)}{32(1-3\lambda)},\;
\xi = \frac{\kappa^{2}\mu^{2}\Lambda}{8(1-3\lambda)},\nb\\
\sigma &=& - \frac{3\kappa^{2}\mu^{2}\Lambda^{2}}{8(1-3\lambda)},
\eqn
where $\kappa^{2} \left[\equiv 8\pi G/c^{4}\right]$ and $\Lambda$ are, 
the Einstein coupling   and cosmological constants, respectively, and 
$\lambda,\; \mu$ and $\omega$ are the three independent coupling constants of the
theory. As pointed out in \cite{LMP}, one can make an analytical continuation of
the parameters $\mu$ and $\omega^{2}$ by
\bq
\lb{2.8}
\mu \rightarrow i \mu,\;\;\;
\omega^{2} \rightarrow - i \omega^{2},
\eq
so that the coupling constants change as,
\bqn
\lb{2.9}
\alpha &\rightarrow& \alpha, \; 
\beta \rightarrow -  \beta,\;
\gamma \rightarrow  - \gamma ,\;
\zeta \rightarrow  -  \zeta,\nb\\
\eta &\rightarrow& - \eta,\;
\xi \rightarrow - \xi,\;
\sigma \rightarrow - \sigma.
\eqn

In this paper, we shall not impose the  ``detailed-balance" conditions given by
Eq.(\ref{2.7}), so that all the constants appearing in the Lagrangian densities 
given by Eq.(\ref{2.5}) are independent and otherwise arbitrary, subject to the
constraint,
\bq
\lb{2.10}
\alpha\left(3\lambda - 1\right) > 0,
\eq
a condition that will be clear when we study cosmological models in the next
section.

In the IR limit, all the quadratic terms of $R_{ij}$ are dropped out, and the 
total action  reduced to
\bqn
\lb{2.11}
S_{total} &\simeq&   \int{dt dx^{i} N \sqrt{g}\left[\alpha\left(K_{ij}K^{ij} 
- \lambda K^{2}\right)\right.}\nb\\
& & \left. +  \eta R^{2} + \xi R + \sigma\right],
\eqn
which will reduce to the Einstein-Hilbert action,
\bq
\lb{2.12}
S_{EH} = \frac{1}{16\pi G}\int{d^{4}x\sqrt{\tilde{g}} 
\left(\tilde{R}_{4}\left[\tilde{g}\right]   - 2\Lambda_{EH}\right)},
\eq
by setting $x^{0} = ct$,
\bqn
\lb{2.13}
\lambda &=& 1, \;\;\; c = \sqrt{\frac{\xi}{\alpha}},\nb\\
16\pi G &=&  \sqrt{\frac{\xi}{\alpha^{3}}}, \;\;\; 
\Lambda_{EH} = - \frac{\sigma}{2\alpha}, 
\eqn
where 
\bqn
\lb{2.14}
\tilde{g}_{00}   &=& -N^{2} + g^{ij}N_{i}N_{j},\;\;
\tilde{g}_{0i}   = N_{i},\nb\\
\tilde{g}_{ij}   &=& g_{ij},\;\;
\sqrt{\tilde{g}} = N\sqrt{{g}}. 
\eqn
Note that Condition (\ref{2.10}), together with the one that $c$ is real, requires $\Lambda < 0$. To get
a positive $\Lambda$, one can invoke the analytical continuation of
the parameters $\mu$ and $\omega^{2}$, given by Eqs.(\ref{2.8}) and (\ref{2.9}).

\section{Cosmological Models in the Horava-Lifshitz theory}

\renewcommand{\theequation}{3.\arabic{equation}} \setcounter{equation}{0}

The homogeneous and isotropic universe is described by the metric,
\bq
\lb{3.1}
ds^{2} = - dt^{2} + a^{2}(t)\left(\frac{dr^{2}}{1 - kr^{2}}
+ r^{2}d^{2}\Omega\right),
\eq
where $d^{2}\Omega
\equiv d\theta^{2} + \sin^{2}\theta d\phi^{2}$, and  $k = 0, \pm 1$.
For a perfect fluid, 
\bq
\lb{3.2}
T_{ab} = \left(\rho + p\right) u_{a}u_{b} + p g_{ab},
\eq
where $u_{a} = \delta^{t}_{a}$ denotes the four-velocity of the fluid,
the field equations of the Horava-Lifshitz theory can be costed in the 
forms,
\bqn
\lb{3.3a}
3\alpha\left(3\lambda - 1\right)H^{2} &=& \rho - \sigma - \frac{6k\xi}{a^{2}}\nb\\
& & - \frac{12k^{2}\left(\zeta + 2\eta\right)}{a^{4}},  \\
\lb{3.3b}
\dot{\rho} + 3H\left(\rho + p\right) &=&0,
\eqn
where $H = \dot{a}/{a}$. 

Introducing the following quantities,
\bqn
\lb{3.3c}
8\pi G &=& \frac{c^{4}}{\alpha(3\lambda -1)},\;\;\;
\rho_{\Lambda} = - p_{\Lambda} = - \sigma,\nb\\
\rho_{k} &=&  -3p_{k}  \equiv \frac{\rho_{k}^{(0)}}{a^{2}},\;
\rho_{dr} =  3p_{dr} \equiv \frac{\rho_{dr}^{(0)}}{a^{4}},
\eqn
where
\bqn
\lb{3.3d}
\rho_{k}^{(0)} &\equiv& \frac{3k}{4(3\lambda -1)} \left(\kappa^{2}\mu^{2}\Lambda
             + 4\alpha(3\lambda -1)^{2}\right),\nb\\
\rho_{dr}^{(0)} &\equiv& - 12 k^{2}(\zeta + 3\eta) = -  \frac{3\kappa^{2}\mu^{2}}{8(3\lambda -1)}k^{2},
\eqn
we find that Eqs.(\ref{3.3a}) and (\ref{3.3b}) can be written in the form,
\bqn
\lb{3.3e}
& & H^{2} + \frac{k}{a^{2}} = \frac{8\pi G}{3c^{4}} \rho_{t},  \\
\lb{3.3f}
& & \dot{\rho}_{i} + 3H\left(\rho_{i} + p_{i}\right) = 0,
\eqn
where ${\rho}_{i} \equiv \left(\rho, \; \rho_{\Lambda},\; \rho_{k},\; \rho_{dr}\right)$, and 
\bq
\lb{3.3g}
\rho_{t}  \equiv  \sum_{i}{\rho_{i}}, \;\;\;  p_{t}  \equiv \sum_{i}{p_{i}}.
\eq
Eqs.(\ref{3.3e}) and (\ref{3.3f}) take exactly the forms of those given in Einstein's theory of gravity.
It is interesting to note that the  term $\rho_{dr}$ also appears  in the brane world scenarios
\cite{branes,KK}. 

\section{Thermodynamics of the Cosmological Models in the Horava-Lifshitz theory}

\renewcommand{\theequation}{4.\arabic{equation}} \setcounter{equation}{0}

Thermodynamics in cosmology has been extensively studied either in Einstein's theory of gravity
\cite{Gong07a} or in modified theories of gravity, such as brane worlds \cite{Cai07}. In this section,
we shall generalize such studies to the HL cosmology. 

The apparent horizon for the FRW model is defined as \cite{Hay02,Wang03}
\bq
\lb{4.1}
f \equiv g^{ab}\tilde{r}_{,a} \tilde{r}_{,b} = 1 - \left(H^{2} + \frac{k}{a^{2}}\right)\tilde{r}^{2} = 0,
\eq
from which we find that
\bq
\lb{4.2}
\tilde{r}_{A}(t) = \frac{1}{\sqrt{H^{2} + \frac{k}{a^{2}}}},
\eq
where $\tilde{r } \equiv a(t) r$ denotes the geometric radius of the two spheres $t, \;r  = Constants$.
Following \cite{CK05}, we define the horizon temperature and entropy as,
\bqn
\lb{4.3}
T_{A} &\equiv & \frac{1}{2\pi \tilde{r}_{A}} = \frac{\sqrt{H^{2} + \frac{k}{a^{2}}}}{2\pi},\nb\\
S_{A} &\equiv&  \frac{\pi \tilde{r}_{A}^{2}}{G} = \frac{\pi}{G\left(H^{2} + \frac{k}{a^{2}}\right)}.
\eqn
Introducing the mass-like function $M(t,r)$ \cite{Gong07b} and the normal vector $k^{a}$ to the 
horizon by, 
\bqn
\lb{4.4}
M(t,r) &=& \frac{\tilde{r}}{2G}\left(1 -  g^{ab}\tilde{r}_{,a} \tilde{r}_{,b} \right)\nb\\
&=& \frac{\tilde{r}}{2G}\left(2  -   \left(H^{2} + \frac{k}{a^{2}}\right)\tilde{r}^{2}\right),\nb\\
k^{a} &=& \delta^{a}_{t} - rH\delta^{a}_{r},
\eqn
we find that the energy flow through the horizon is given by
\bqn
\lb{4.5}
dE_{A}  &=& \left. dE(t, r)\right|_{\tilde{r} = \tilde{r}_{A}}
\equiv  \left. k^{a}\nabla_{a} M(t, r)\right|_{\tilde{r} = \tilde{r}_{A}}dt\nb\\
&=& - \frac{H\left(\dot{H} - \frac{k}{a^{2}}\right)}{G\left(\dot{H} + \frac{k}{a^{2}}\right)^{3/2}}.
\eqn
On the other hand, from Eq.(\ref{4.3}) it can be shown that
\bq
\lb{4.6}
T_{A}dS_{A} = - \frac{H\left(\dot{H} - \frac{k}{a^{2}}\right)}{G\left(\dot{H} + \frac{k}{a^{2}}\right)^{3/2}},
\eq
that is, the first law, $T_{A}dS_{A} = dE_{A}$, of thermodynamics holds on the apparent horizon.
 It should be noted that
the above considerations are purely geometric, and were not involved with any field equations.
Therefore, they hold for any metric theories, as mentioned in \cite{Gong07a}.

In the rest of this section, we consider the  first law of thermodynamics outside the apparent 
horizon, which can be written as \cite{Gong07a},
\bq
\lb{4.7}
T dS(t,V) = Td(sV) = d\left(\rho(T) V\right) + p(T)dV,
\eq
where $S(T,V) = s(T,V)V$ denotes the total entropy of the system,  $s(T,V)$ the entropy 
density, and $T$ the temperature. Here we consider that $T$ and $V$ are two independent 
variables. In other words,  we consider a region of the universe with a finite radius $r$.

Before proceeding further, we would like to note that, in \cite{Gong07a}, it was considered the 
case where the system consists the whole region inside the apparent horizon $ \tilde{r} \le \tilde{r}_{A}$,
so that the total volume $V$ of the system depends on $T$. For detail, we refer 
readers to \cite{Gong07a}. In this paper, 
we shall not consider such a possibility. Then, from the condition $\partial^{2}S/\partial T\partial V = 
\partial^{2}S/\partial V\partial T$, we find that 
 \bq
 \lb{4.8}
 \frac{dp}{dT} = \frac{\rho + p}{T}.
 \eq
 Substituting the above into Eq.(\ref{4.7}), we find that 
 \bq
 \lb{4.9}
 d\left(sV - \frac{\rho + p}{T} V\right) = 0,
 \eq
 which has the general solution,
  \bq
 \lb{4.10}
 s = \frac{\rho + p}{T}   + \frac{s_{0}}{V},
 \eq
 where $s_{0}$ is a constant and usually set to zero \cite{Gong07a}. However, here we shall leave this possibility
 open. The volume $V$ is given by
 \bqn
 \lb{4.11}
 V &=& \int{\sqrt{g_{3}}d^{3}x} = \frac{4\pi}{3}a^{3} \int_{0}^{r}{\frac{{r'}^{2}}{\sqrt{1 -k {r'}^{2}}} dr'} \nb\\
 &\equiv&  \frac{4\pi}{3}a^{3}V_{0}(r,k).
  \eqn
Inserting Eqs.(\ref{4.8}) - (\ref{4.11}) into Eq.(\ref{4.7}), and considering the conservation law (\ref{3.3f}) for
$\rho_{i} = \rho$ and $p_{i} = p$,  we find that 
\bq
\lb{4.12}
d\ln\left(s - \frac{s_{0}}{V}\right) = - 3\left(\frac{da}{a}\right),
\eq
which has the general solution 
\bq
\lb{4.13}
s = \frac{1}{a^{3}}\left(s_{1} + \frac{3s_{0}}{4\pi V_{0}(r, k)}\right),
\eq
where $s_{1}$ is another integration constant. 
Combining Eq.(\ref{4.13}) with Eq.(\ref{4.10}) we find that 
\bq
\lb{4.14}
\rho + p = \frac{s_{1}}{2\pi a^{3}} \sqrt{H^{2} + \frac{k}{a^{2}}}.
\eq
From Eqs.(\ref{3.3e}) and (\ref{3.3f}), it can be shown that Eq.(\ref{4.14}) can be
further written as
\bq
\lb{4.15}
\dot{H} + \frac{2s_{1}G}{a^{3}}  \sqrt{H^{2} + \frac{k}{a^{2}}} = \frac{\tilde{k}}{a^{2}}
    + \frac{16\pi G}{3}\rho_{dr},
   \eq
where
\bq
\lb{4.16}
\tilde{k} \equiv - \frac{2k\kappa^{2}\mu^{2}\Lambda}{4\alpha(3\lambda -1)^{2}}.
\eq
Clearly, the first law of thermodynamics holds only when condition (\ref{4.15}) is satisfied. 
In other words, it   holds only for the fluid that satisfies the above condition. This is also true
in Einstein's theory of gravity, in which it was shown that the first law of thermodynamics 
requires that the fluid must consist only three parts \cite{Gong07b}, the cosmological constant,
non-relativistic matter, and dark radiation. The three parts are coupled each other  as
\cite{Gong07b},
\bq
\lb{4.17}
\rho = \rho_{\Lambda} + 2\sqrt{\rho_{0} \rho_{\Lambda}} \left(\frac{a_{0}}{a}\right)^{3}
+ \rho_{0}\left(\frac{a_{0}}{a}\right)^{6}.
\eq

\section{Classification of the FRW universe in the Horava-Lifshitz theory}

\renewcommand{\theequation}{5.\arabic{equation}} \setcounter{equation}{0}

Considering  the equation of state given by
\bq
\lb{3.4}
p = w \rho,
\eq
where $w$ is an arbitrary real constant, from Eq.(\ref{3.3b})  we find that,
\bq
\lb{3.5}
\rho = \rho_{0} \left(\frac{a_{0}}{a}\right)^{3(1+w)},
\eq
where $\rho_{0}$ and $a_{0}$ are the integration constants. Since
$\rho_{0}$ represents the energy density when $a = a_{0}$, we shall
assume that it is strictly positive $\rho_{0} > 0$.  Without loss of
generality, we can always set $a_{0} = 1$. Then, it can be shown that
the Friedmann equation  (\ref{3.3a}) can be cast in the form \cite{szy,Ha09},
\bq
\lb{3.6}
\frac{1}{2} {a^{*}}^{2} + V(a) = 0,
\eq
where $a^{*} \equiv da(t)/d(H_{0} t)$, and
\bq
\lb{3.7}
V(a) = - \frac{1}{2}\left(\frac{\Omega_{m}}{a^{1+3w}}
+ \Omega_{k} + \Omega_{\Lambda} a^{2} 
+  \frac{\Omega_{dr}}{a^{2}}\right), 
\eq
with
\bqn
\lb{3.8} 
\Omega_{m} &\equiv& \frac{\rho_{0}}{3\alpha(3\lambda -1)H^{2}_{0}},\;
\Omega_{\Lambda} \equiv - \frac{\sigma}{3\alpha(3\lambda -1)H^{2}_{0}},\nb\\
\Omega_{k} &\equiv& - \frac{2k\xi}{\alpha(3\lambda -1)H^{2}_{0}},\;
\Omega_{dr}  \equiv \frac{4(\zeta + 3\eta) k^{2}}{\alpha(3\lambda -1)H^{2}_{0}}.\;\;\;
\eqn
Thus, the acceleration of the universe is given by
\bq
\lb{acceleration}
a^{**} = - \frac{dV(a)}{da} =  \frac{\ddot{a}}{H^{2}_{0}}.
\eq
As mentioned previously, we shall not impose the ``detailed-balance" conditions,
except the condition given by Eq.(\ref{2.10}), so that the Friedmann equation
(\ref{3.3a}) has the correct coupling sign between the Hubble expansion factor
and the matter fields. Under such an assumption, all the coupling constants 
appearing in the Lagrangian densities (\ref{2.5}) are free parameters, so
that all the quantities  defined in Eq.(\ref{3.8}) can have any signs, except for 
$\Omega_{m}$ for which we assume that it is always positively-defined,
 $\Omega_{m} > 0$.  When $k = 0$, we have $\Omega_{k} = \Omega_{dr} = 0$, and
the corresponding Friedmann equation reduces to that of Einstein's
theory, studied in detail in \cite{Ha09}. Therefore, in the rest of this paper, we 
shall assume that $k \not=0$.  Then, it is found convenient to distinguish the 
three cases: $\Omega_{\Lambda} = 0$, $\Omega_{\Lambda} > 0$ and $\Omega_{\Lambda} 
< 0$. In each of them there are seven sub-cases: 
\bqn
\lb{cases}
& & (i) \; w > \frac{1}{3}; \;\;\; (ii) \; w =  \frac{1}{3};  \;\;\; (iii) - \frac{1}{3} < w < \frac{1}{3};  \nb\\
& & (iv)\; w = - \frac{1}{3};\;\;\; (v) \; - 1 < w < -  \frac{1}{3};   \nb\\
& & (vi) \; w = -1;\;\;\; (vii) \; w < -1. 
\eqn
In the following we shall consider each of them
separately.

\subsection{$\Omega_{\Lambda} = 0$}

When $\Omega_{\Lambda} = 0$, Eq.(\ref{3.7}) reduces to
\bq
\lb{3.9}
V(a) = - \frac{1}{2}\left(\frac{\Omega_{m}}{a^{1+3w}}
+ \Omega_{k}  +  \frac{\Omega_{dr}}{a^{2}}\right), 
\eq
from which we find that
\bq
\lb{3.10}
V'(a) =   \frac{1}{2a^{2+3w}}\left(\left(1+3w\right)\Omega_{m}
+ 2\Omega_{dr}a^{3w-1}\right),
\eq 
where $V'(a) \equiv dV(a)/da$. 

\subsubsection{$ w > \frac{1}{3}$}

In this case, from Eq.(\ref{3.10}) we can see that when $\Omega_{dr} < 0$, the potential 
has a maximum at 
\bq
\lb{3.11}
a_{max} = \left(\frac{(1+3w)\Omega_{m}}{2\left|\Omega_{dr}\right|}\right)^{1/(3w-1)},
\eq
where $V'\left(a_{max}\right) = 0$. When  $\Omega_{dr} > 0$, such a point does not exist,
and $V(a)$ is a monotonically increasing function. Therefore, we shall consider the two cases 
$\Omega_{dr} > 0$ and $\Omega_{dr} < 0§$ separately.

{\bf Case A.1.1) $\; \Omega_{dr} < 0$}: In this case, we have 
\bq
\lb{3.12}
V(a) = \cases{- \infty,& $ a = 0$,\cr
-\Omega_{k}/2, & $a \rightarrow \infty$.\cr}
\eq
Thus, depending on the signs of $\Omega_{k}$, the potential has different behaviors. 

\begin{figure}
\includegraphics[width=\columnwidth]{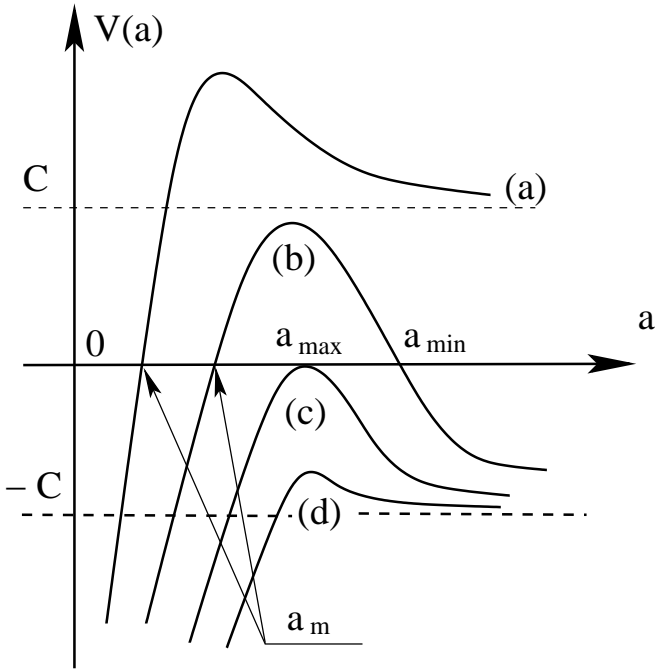}
\caption{The potential given by Eq.(\ref{3.9})   for  $\Omega_{\Lambda} =   0,
\; w > 1/3$ and $\Omega_{dr} < 0$. (a) $\; \Omega_{k} < 0$;
(b) $\; \Omega_{k} >  0,\;  \Omega_{dr} <  \Omega_{dr}^{c}$;  (c) $\; 
\Omega_{k} >  0,\;  \Omega_{dr} =  \Omega_{dr}^{c}$;
and (d) $\; \Omega_{k} >  0,\;  \Omega_{dr} >  \Omega_{dr}^{c}$, where 
$C \equiv \left|\Omega_{k}\right|/2$.
}
\label{fig1}
\end{figure}

\begin{figure}
\includegraphics[width=\columnwidth]{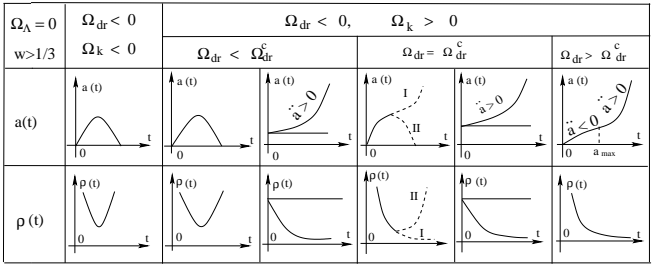}
\caption{The evolution of the universe  with the potential given by Eq.(\ref{3.9})  for  $\Omega_{\Lambda} =   0,\; w > 1/3$ 
and $\Omega_{dr} < 0$. A big bang singularity happens whenever $a(0) = 0$, 
while a big crunch singularity happens
whenever $a(t_{s}) = 0$.}
\label{fig2}
\end{figure}

{\bf Case A.1.1.a) $\; \Omega_{dr} < 0,\; \Omega_{k} < 0$}:  Then,  the potential  
is given by Curve (a) in Fig.\ref{fig1},
from which we can see that there exists
a point $a= a_{m}$ at which we have $V(a_{m}) = 0$. Thus, in this case if the universe starts to
expand at the big bang $a(0) = 0$, it will expand with $\ddot{a} < 0$ until $a = a_{m}$, at which we 
have $\dot{a} = 0$, but we still have $\ddot{a} = - H^{2}_{0}dV(a)/da < 0$. So afterwards,  
the universe will start to
 collapse, until it reaches the point $a(t_{s}) = 0$ again, whereby a big crunch singularity is developed. 
 The evolution of the universe is shown schematically in   Fig. \ref{fig2}.
 
{\bf Case A.1.1.b) $\; \Omega_{dr} < 0,\; \Omega_{k} > 0$}: In this case, 
there exists a critical value $\Omega_{dr}^{c} < 0$ for 
any given $\Omega_{m}$  and $\Omega_{k}$,  which satisfies the conditions
 \bqn
 \lb{3.13}
 V\left(a_{max}, \; \Omega_{m},\; \Omega_{k}, \; \Omega_{dr}^{c} \right) &=& 0,\nb\\
 V'\left(a_{max}, \; \Omega_{m},\; \Omega_{k}, \; \Omega_{dr}^{c}\right) &=& 0,
 \eqn
 as shown in Fig. \ref{fig1}.

\begin{figure}
\includegraphics[width=\columnwidth]{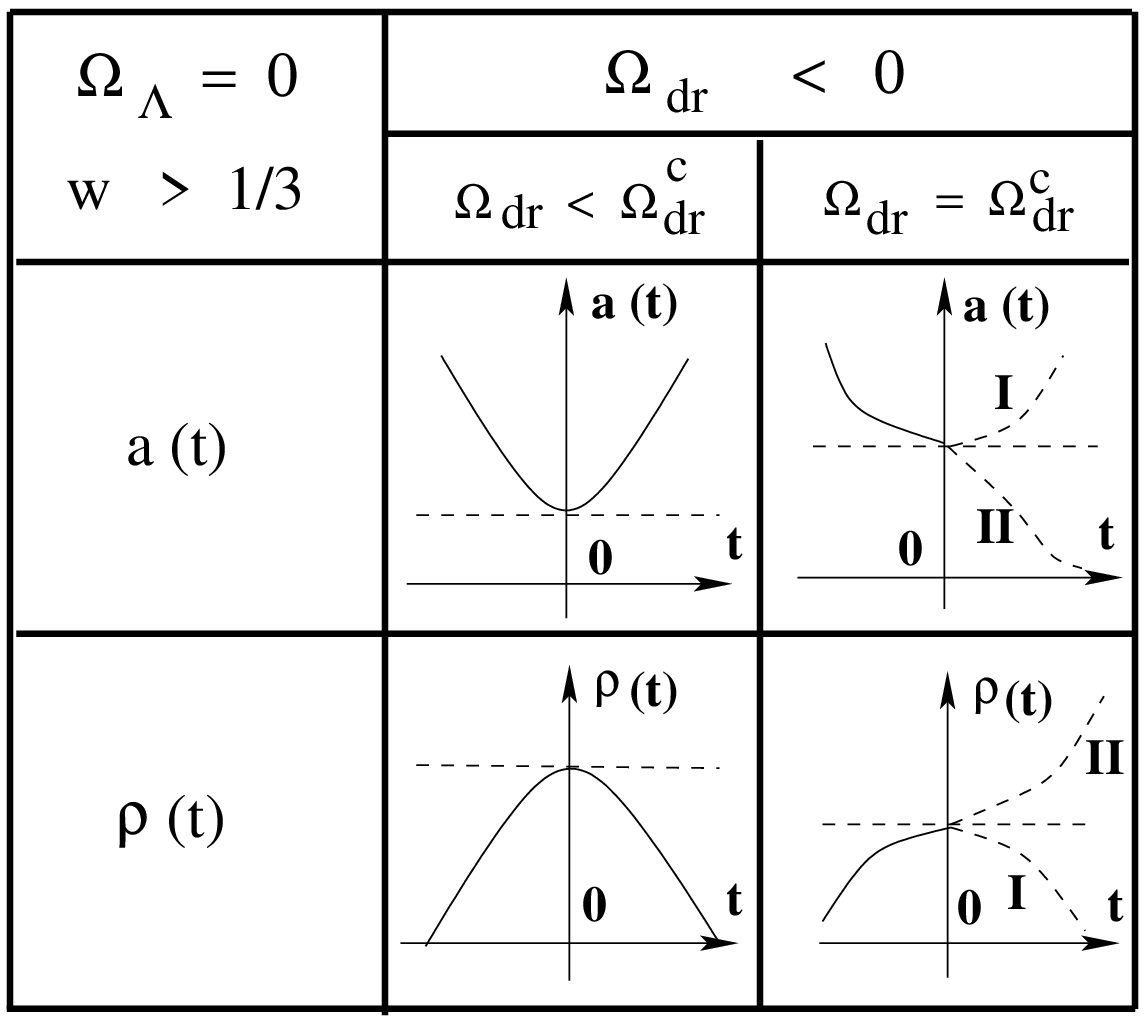}
\caption{The evolution of the universe  with the potential given by Eq.(\ref{3.9})  for  $\Omega_{\Lambda} =   0,\; w > 1/3$ 
and $\Omega_{dr} < 0$. A bouncing universe is allowed in either of these two cases.  }
\label{fig2a}
\end{figure}

 When $\Omega_{dr} <  \Omega_{dr}^{c} < 0$, the potential is given by Curve
 (b), from which we can see that now $V(a) = 0$ has two positive roots, $a_{m}$ 
 and $a_{min}$, where $a_{m} > a_{min}$. In this case, the evolution of the 
 universe depends on its initial condition. If it starts to expand at the big bang, 
 it will expand until $a = a_{m}$ and then  collapse to $a =0$ within
 finite time, whereby a big crunch singularity is developed. This is similar 
 to the last case. However,  if the universe starts to expand at $a_{i} \ge a_{min}$,
  it will expand  forever with a positive acceleration  $\ddot{a} = - H^{2}_{0}dV(a)/da > 0$.
 It is interesting to note  that in the latter case a bouncing universe is also allowed.
 For example, if the universe is initially collapsing at $a_{i} > a_{min}$ with $\dot{a}(t_{i})
 < 0$, then the universe will collapse until $a = a_{min}$. Once it reaches the point $a_{min}$,
where we have $\dot{a}(t_{min}) = 0$ and $ \ddot{a}(t_{min}) > 0$, then the universe will turn 
 around, and starts to expand acceleratingly without further turning-back, as shown 
 by Fig.\ref{fig2a}.
  
 When $\Omega_{dr} =  \Omega_{dr}^{c}$, the potential is given by the Curve
 (c), from which we can see that now $V(a) = 0$ has two degenerate roots, 
 $a_{max} = a_{m} = a_{min} > 0$.  If it
 starts to expand at the big bang, it will expand until $a = a_{max}$. 
 Since we have $V\left(a_{max}\right)
 = 0 = V'\left(a_{max}\right)$, now $a = a_{max}$ represents a stationary 
 point. But, it is not stable, and
 with a small perturbation, it will either collapse to form a big crunch 
 singularity at $a = 0$ or expand
 forever with $\ddot{a} > 0$.   If the universe starts to expand at $a_{i} 
 \ge a_{max}$, it will expand forever
  with a positive acceleration  $\ddot{a} = -H^{2}_{0} dV(a)/da > 0$, as 
  shown by Fig. \ref{fig2}. If it starts to collapse at $a_{i} 
 \ge a_{max}$, the universe will reach the point $a_{max}$ within a finite proper
 time, and afterwards it will stay there forever. However, since now $a = a_{max}$ is not
 a stable point, with a small perturbation, it will either start to expand forever or collapse
 until a big crunch singularity is formed at $a(t_{s}) = 0$, as shown by Fig.\ref{fig2a}.
  
   When $\Omega_{dr}  >  \Omega_{dr}^{c}$, the potential is always negative, 
   and represented by Curve (d).
   Now the universe will start  to expand from a big bang singularity at $a =0$
    forever. But, when $a < a_{max}$ it is decelerating,
   while when $a > a_{max}$ it is accelerating.

\begin{figure}
\includegraphics[width=\columnwidth]{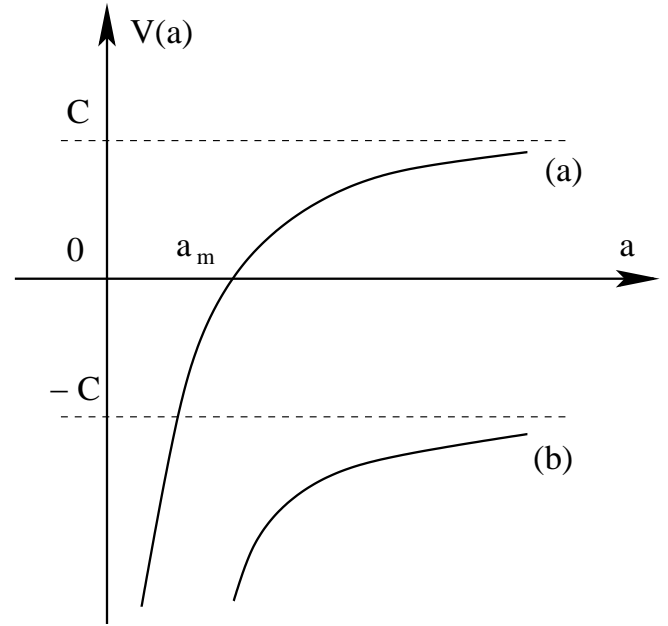}
\caption{The potential given by Eq.(\ref{3.9})   for  $\Omega_{\Lambda} =   0,
\; w > 1/3$ and $\Omega_{dr} > 0$.  (a) $\; \Omega_{k} < 0$;
and (b) $\; \Omega_{k} >  0$, where $C \equiv \left|\Omega_{k}\right|/2$.
}
\label{fig3}
\end{figure}

{\bf Case A.1.2) $\; \Omega_{dr} > 0$}: In this case, 
we have $V'(a) < 0$, and as a result,  the universe is always decelerating, as 
$a(t)$ is increasing. The potential is given by Fig.\ref{fig3}. 

\begin{figure}
\includegraphics[width=\columnwidth]{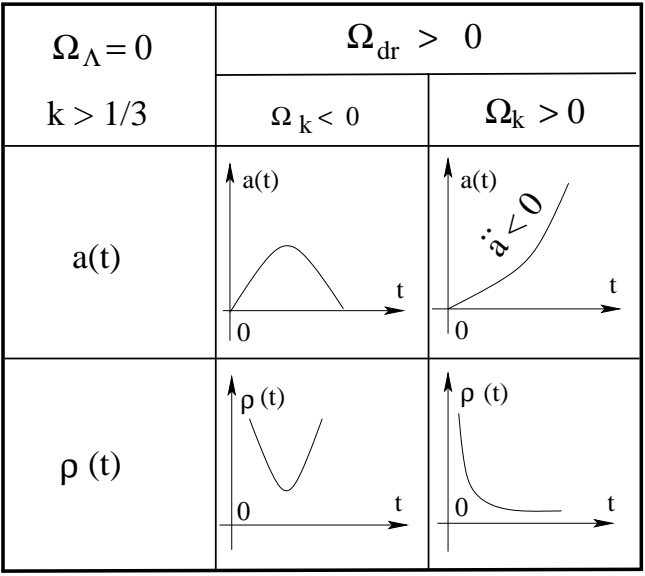}
\caption{The evolution of the universe with the potential given by Eq.(\ref{3.9})  for  $\Omega_{\Lambda} =   0,\; w > 1/3$ 
and $\Omega_{dr} > 0$. A big bang singularity happens at $a(0) = 0$, 
while a big crunch singularity happens at $a(t_{s}) = 0$.}
\label{fig4}
\end{figure}

When $\Omega_{k} <0$, the potential is given by Curve (a) in Fig. \ref{fig3}, 
from which we can see that there exists a point $a_{m}$,
 for which we have $V(a<a_{m}) < 0$, where $V(a_{m}) = 0$. The universe in this 
 case starts to expand from a big bang singularity 
 at $a(0) = 0$ with $\ddot{a} < 0$ until it reaches  its maximal radius $a_{m}$. Afterwards,  
 it starts to collapse
until $a(t_{s}) = 0$ reaches again, whereby a big crunch is developed, as shown by 
Fig. \ref{fig4}.

When $\Omega_{k} > 0$, the potential is given by Curve (b) in Fig. \ref{fig3}, 
from which we can see that the potential is always
negative, and $V'(a) > 0$. Thus, in this case the universe is expanding  from 
a big bang singularity 
 at $a(0) = 0$ forever. It is always decelerating, as $\ddot{a} \propto - 
 dV(a)/da < 0$, as shown by Fig. \ref{fig4}.

\subsubsection{$ w = \frac{1}{3}$}

In this case,  Eq.(\ref{3.9})  reduces to
\bq
\lb{3.13a}
V(a) = - \frac{1}{2}\left( \Omega_{k}    +  \frac{\Omega_{\delta}}{a^{2}}\right), 
\eq
where $\Omega_{\delta} \equiv  \Omega_{m} + \Omega_{dr}$. 

When $\Omega_{\delta} > 0$ and  $\Omega_{k} < 0$, the potential is given by Curve
(a) in Fig. \ref{fig3}, and the corresponding motion of the universe is similar to the case
of $\Omega_{\Lambda} = 0,\; w > 1/3, \Omega_{dr} > 0$ and $\Omega_{k} < 0$.
In particular,   there exists a maximal radius $a_{m}$. When $a < a_{m}$ we have 
$V(a) < 0$ and $V'(a) > 0$. Thus, the universe 
in this case starts to expand from the 
big bang at  $a(0) =0$ with $\ddot{a} < 0$ until its maximal radius $a_{m}$. 
Afterwards, it will start to collapse until
the moment where $a(t_{s}) = 0$ again, at which a big crunch singularity is formed, 
as shown by the first case in Fig. \ref{fig4}.

When $\Omega_{\delta} > 0$ and $\Omega_{k} > 0$, the potential is given by Curve (b) in Fig. \ref{fig3}, 
and the motion of the universe is similar to the case of  $\Omega_{\Lambda} = 0,\; w > 1/3, \Omega_{dr} > 0$ 
and $\Omega_{k} > 0$, given by the second case in Fig. \ref{fig4}.

\begin{figure}
\includegraphics[width=\columnwidth]{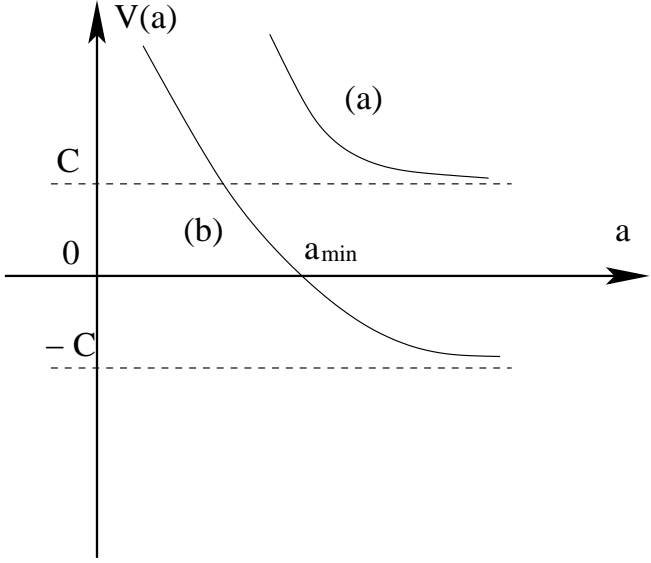}
\caption{The potential given by Eq.(\ref{3.13a})   for  $\Omega_{\Lambda} =   0$ 
and $w = 1/3$:        (a) $\; 
 \Omega_{\delta} < 0,\; \Omega_{k} <  0$;
and   (b) $\; \Omega_{\delta} < 0,\; \Omega_{k} >  0$, where $C \equiv 
\left|\Omega_{k}\right|/2$.
}
\label{fig5}
\end{figure}

 When $\Omega_{\delta} < 0$, from Eq.(\ref{3.13a}) we can see that the potential 
 is always positive for $\Omega_{k}  < 0$, given by Curve (a) in Fig. \ref{fig5}. As a result, the motion in this
  case is forbidden. 
  
\begin{figure}
\includegraphics[width=\columnwidth]{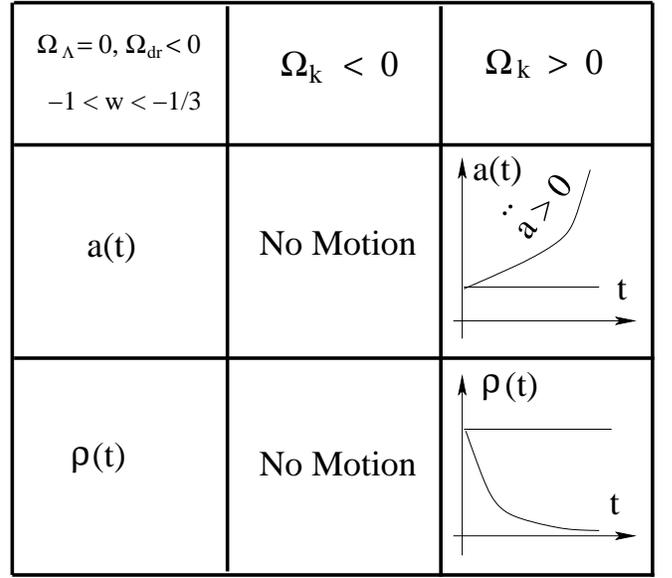}
\caption{The evolution of the universe with the potential given by Eq.(\ref{3.13a}) for  $\Omega_{\Lambda} =   0$,
and $w = 1/3$. A big bang singularity happens at $a(0) = 0$.}
\label{fig6}
\end{figure}

  When   $\Omega_{\delta} < 0$ and $\Omega_{k}  >0$, there exists a point $a_{mim}$ for which we have $V(a > a_{mim}) < 0$, as 
 shown by Curve (b) in Fig.\ref{fig5}. Then, the universe in this case will expand from a non-zero radius, say,  
 $a=a_{i} \ge a_{m}$ until $a = \infty$  with $\ddot{a} > 0$, as shown in Fig. \ref{fig6}. In this case, a bouncing universe
 is also allowed, if it starts to collapse initially at $a_{i} > a_{min}$ with $\dot{a}(t_{i}) < 0$. Then, it will collapse until
 it reaches $a = a_{min}$. Afterwards, it will expand forever with $\ddot{a} > 0$, as shown by the first case in Fig.\ref{fig2a}.

\subsubsection{$ - \frac{1}{3} < w < \frac{1}{3}$}

In this case,   Eq.(\ref{3.9}) reduces to  
\bq
\lb{3.14}
V(a) = - \frac{1}{2}\left(\Omega_{k} + \frac{1}{a^{2}}\left(\Omega_{dr}
+ \Omega_{m}a^{1-3w}\right)\right).
\eq
Thus, depending on the signs of $\Omega_{dr}$, the potential can have different 
properties. 

\begin{figure}
\includegraphics[width=\columnwidth]{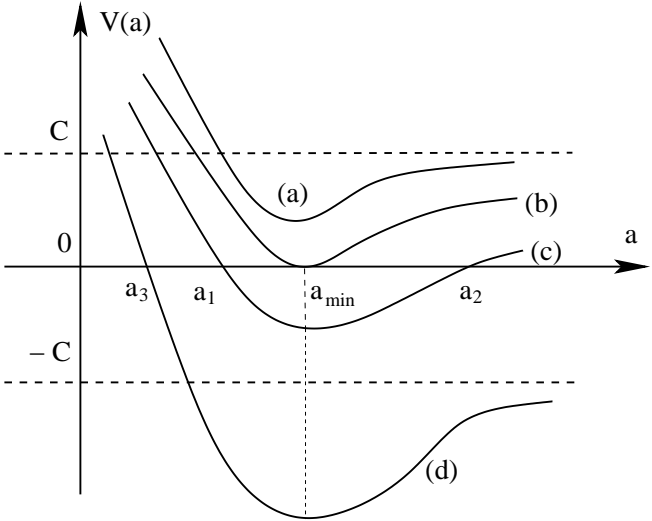}
\caption{The potential given by Eq.(\ref{3.14})   for  $\Omega_{\Lambda} =   0,\;
\Omega_{dr} < 0$  and $- 1/3 < w < 1/3$:   
(a) $\; \Omega_{k} < 0,\; \left|\Omega_{dr}\right| > \left|\Omega_{dr}^{c}\right|$;
(b) $\; \Omega_{k} < 0,\; \left|\Omega_{dr}\right| = \left|\Omega_{dr}^{c}\right|$;
(c) $\; \Omega_{k} < 0,\; \left|\Omega_{dr}\right| < \left|\Omega_{dr}^{c}\right|$;
and
(d) $\; \Omega_{k} > 0$,  where $C \equiv 
\left|\Omega_{k}\right|/2$ and $\left|\Omega_{dr}^{c}\right|$ is defined by 
Eq.(\ref{3.18}).
}
\label{fig7}
\end{figure}

{\bf Case A.3.1) $\; \Omega_{dr} < 0$}: Then,  we have 
\bq
\lb{3.15}
V(a) = \cases{ \infty, & $a = 0$,\cr
-\Omega_{k}/2,& $a = \infty$,\cr}
\eq
and 
\bq
\lb{3.16}
V'(a) =   - \frac{1}{a^{3}}\left(\left|\Omega_{dr}\right| 
- \frac{(1+3w)\Omega_{m}}{2} a^{1-3w}\right).
\eq 
Therefore, in the present case the potential always has a minimum at
\bq
\lb{3.17}
a_{min}  = \left(\frac{2\left|\Omega_{dr}\right|}
{(1+3w)\Omega_{m}}\right)^{\frac{1}{1-3w}},
\eq
as shown in Fig. \ref{fig7}. 

\begin{figure}
\includegraphics[width=\columnwidth]{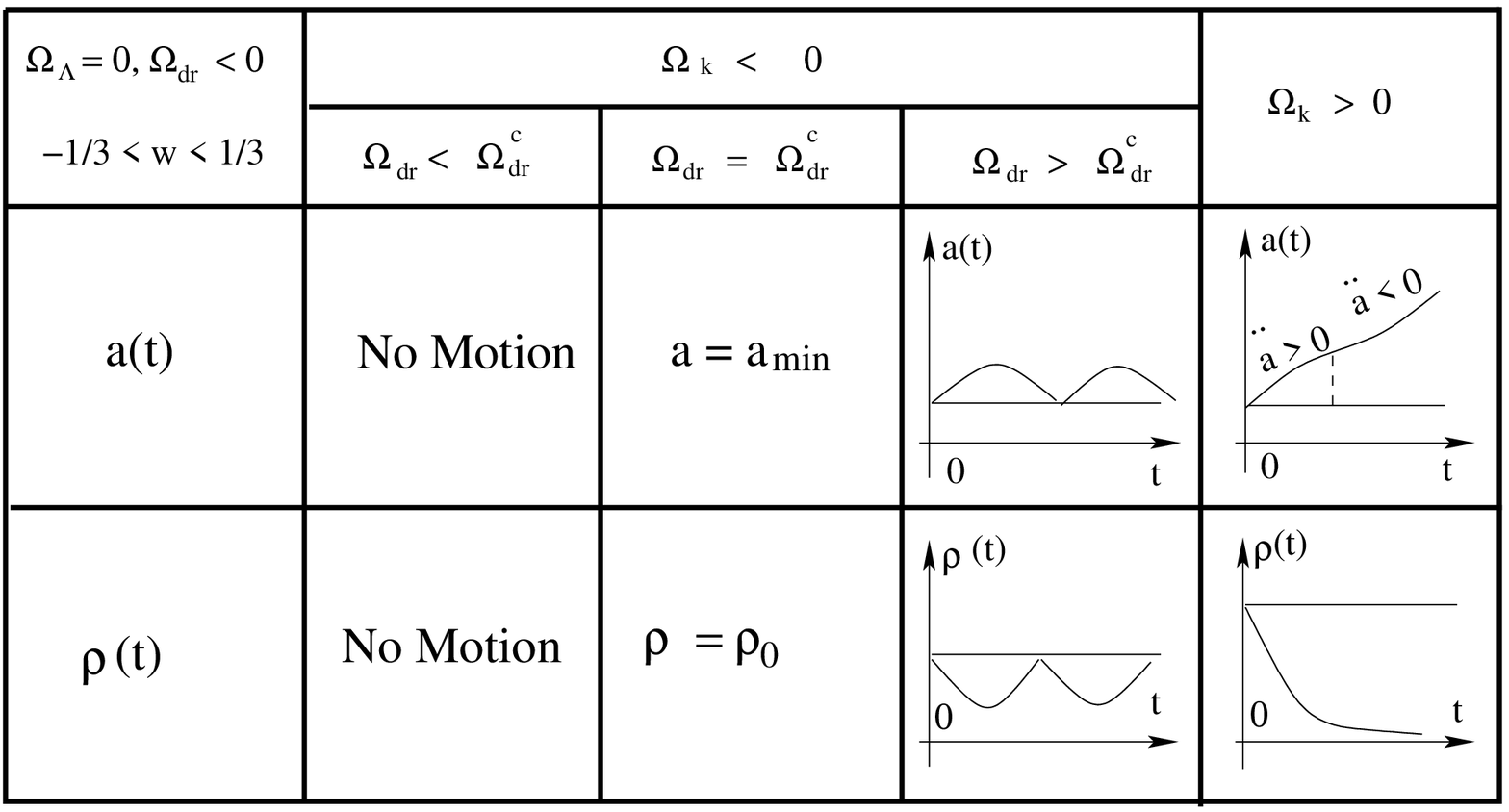}
\caption{The evolution of the universe with the potential given by Eq.(\ref{3.14})   for  $\Omega_{\Lambda} =   0,\;
\Omega_{dr} < 0$  and $- 1/3 < w < 1/3$. }
\label{fig8}
\end{figure}

When $\Omega_{k} < 0$, there exists a critical value, $\Omega_{dr}^{c}$ given by
\bq
\lb{3.18}
\left|\Omega_{dr}^{c}\right| = \frac{(1+3w)a^{2}_{min}}{1-3w}\left|\Omega_{k}\right|,
\eq
so that when $\left|\Omega_{dr}\right| > \left|\Omega_{dr}^{c}\right|$ the
potential $V(a)$ is always positive, as shown by Curve (a), and the motion is forbidden.
When $\left|\Omega_{dr}\right| = \left|\Omega_{dr}^{c}\right|$ the
potential $V(a)$ is always positive, except  the point $a = a_{min}$, at which we
have $V(a_{min},\Omega^{c}_{dr}) = 0 = V'(a_{min},\Omega^{c}_{dr})$, as shown by Curve
(b). This point represents a static universe. In contrast to Einstein's static 
universe in General Relativity \cite{d'Inverno}, this static universe  seems stable, 
as now it corresponds to a minimum of the potential. When $\left|\Omega_{dr}\right| < 
\left|\Omega_{dr}^{c}\right|$ the potential $V(a)$ is non-positive only for 
$a \in [a_{1}, \; a_{2}$], where $a_{1,2}$ are the two positive roots of $V(a) = 0$
with $a_{2} > a_{1}$. Then, the universe   is oscillating between the two
radii $a_{1}$ and $a_{2}$ without forming any kind of spacetime singularities.
During the period $a \in[a_{1}, a_{min})$ the universe is accelerating, while during
the period $a \in(a_{min}, a_{2})$ it is decelerating, as shown in Fig. \ref{fig8}.
Therefore, in this case we have a bouncing cyclic universe. 

When $\Omega_{k} > 0$, the potential is negative only for $a > a_{3}$ where $a_{3}$ is the 
real and positive root of $V(a) = 0$, as shown by Curve (d) in Fig.\ref{fig7}.
Therefore, in this case the universe starts to expand from a non-zero and finite radius,
say, $a_{i} \ge a_{3}$, and shall expand forever. During the period
$ a_{3} \ge a < a_{min}$, it is accelerating, while during the period
$ a > a_{min}$, it is decelerating. But the universe never stops expanding until
$a = \infty$, as shown in Fig.\ref{fig8}. A bouncing universe is also allowed, if it 
starts to collapse at $a_{i} > a_{3}$ with $\dot{a}(t_{i}) < 0$. Then, the corresponding motion
is similar to that given by the first case in Fig.\ref{fig2a}.

{\bf Case A.3.2) $\; \Omega_{dr} > 0$}: In this case we have 
\bq
\lb{3.19}
V(a) = \cases{ -\infty, & $a = 0$,\cr
-\Omega_{k}/2,& $a = \infty$,\cr}
\eq
and 
\bq
\lb{3.20}
V'(a) =    \frac{1}{a^{3}}\left(\left|\Omega_{dr}\right| +
 \frac{(1+3w)\Omega_{m}}{2} a^{1-3w}\right) \ge 0,
\eq 
where equality holds only when $a = \infty$.

When $\Omega_{k} < 0$, the potential has the same form as that given by Curve (a) in Fig.\ref{fig3}  for the case $\Omega_{\Lambda} = 0,\;
w > 1/3, \Omega_{dr} > 0$ and $\Omega_{k} < 0$. As a result, the motion of the universe is also similar to that case.

When  $\Omega_{k} > 0$, the potential is given by Curve (b) in Fig. \ref{fig3}  for the case $\Omega_{\Lambda} = 0,\;
w > 1/3, \Omega_{dr} > 0$ and $\Omega_{k} > 0$. Therefore, in this case the motion of the universe can be immediately deduced from
there, and given as that described in Fig.\ref{fig4} for  $\Omega_{k} > 0$.
 
 It should be noted that although in these two cases the motion of the universe has similar characteristics, the detail could be different, and
 when one fits the models to observational data, one can get completely different conclusions. Since in this paper we do not consider the fitting,
 we shall not distinguish them here.



\subsubsection{$ w = - \frac{1}{3} $}

In this case,   Eq.(\ref{3.9}) reduces to  
\bq
\lb{3.21}
V(a) = - \frac{1}{2}\left(\Omega_{\Delta} + \frac{\Omega_{dr}}{a^{2}}\right),
\eq
where $\Omega_{\Delta} \equiv \Omega_{k} + \Omega_{m}$. 

When $\Omega_{dr} > 0$ and $\Omega_{k} < 0$, the potential  is given by Curve (a) in Fig.\ref{fig3}.
Therefore, the motion of the universe is similar to that corresponding case. 
 
When $\Omega_{dr} > 0$ and $\Omega_{k} > 0$, the potential is always negative and asymptotically
approaches to $- \Omega_{k}/2$, as shown by Curve (b) in Fig. \ref{fig3}.  Clearly, in this case the
universe starts to expand from the big bang singularity at $a(0) = 0$ and then expands forever
with $\ddot{a} < 0$.

When $\Omega_{dr} < 0$ and $\Omega_{k} < 0$, the potential is always positive, as shown by 
Curve (a) in Fig.\ref{fig5}. So,  the motion is forbidden. 

When $\Omega_{dr} < 0$ and $\Omega_{k} > 0$, the potential is  negative only when $a \ge a_{mim}$,
as shown by Curve (b) in Fig.\ref{fig5}.  Therefore, in the present case the 
universe will start to expand from a finite radius, say, $a_{i} \ge a_{m}$, and shall expand forever.
Since now we have $dV(a)/da < 0$, we can see that in this case the universe is always accelerating.
Note that now no spacetime singularity exists, as we always have $a \ge a_{mim} > 0$.
Certainly, in this case a bouncing universe is also allowed.

\subsubsection{$ -1 < w < -  \frac{1}{3}$}

In this case,   Eq.(\ref{3.9}) can be written as
\bq
\lb{3.22}
V(a) = - \frac{1}{2}\left(\Omega_{k} + \frac{\Omega_{dr}}{a^{2}}
+ \Omega_{m}a^{3|w| -1}\right),
\eq
from which we have
\bq
\lb{3.23}
V'(a) = - \frac{1}{2a^{3}}\left(\left(3|w| -1\right)\Omega_{m}a^{3|w| +1}  - 2 \Omega_{dr}\right).
\eq

\begin{figure}
\includegraphics[width=\columnwidth]{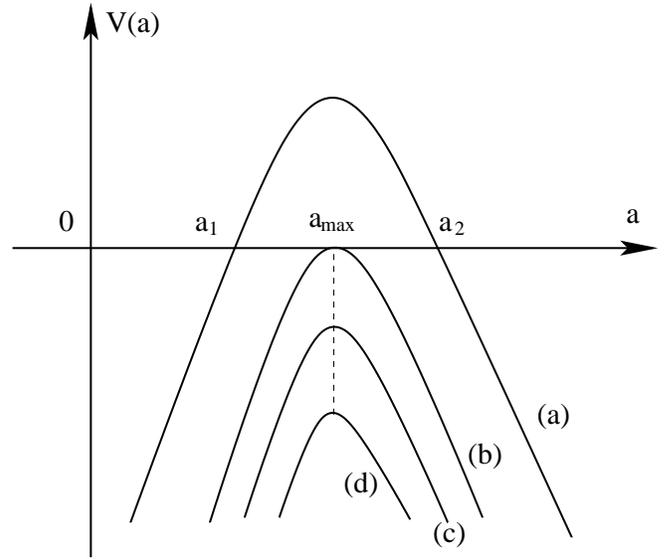}
\caption{The potential given by Eq.(\ref{3.22})   for  $\Omega_{\Lambda} =   0, \;  \Omega_{dr} > 0 $  and $- 1 < w < - 1/3$:   
(a) $\; \Omega_{dr}  <  \Omega_{dr}^{c}, \;  \Omega_{k} < 0$;   (b) $\; \Omega_{dr}  =  \Omega_{dr}^{c}, \;  \Omega_{k} > 0$; 
(c) $\;  \Omega_{dr}  >  \Omega_{dr}^{c}, \;  \Omega_{k} < 0$;  and (d) $\; \Omega_{dr} > 0, \;  \Omega_{k} > 0$,
 where   $\Omega_{dr}^{c}$ is defined by Eq.(\ref{3.26}).
}
\label{fig9}
\end{figure}

 {\bf Case A.5.1) $\; \Omega_{dr} > 0$}: In this case we have 
\bq
\lb{3.24}
V(a) = \cases{ -\infty, & $a = 0$,\cr
- \infty,& $a = \infty$,\cr}
\eq
and the potential has a maximum at 
\bq
\lb{3.25}
a_{max}  = \left(\frac{2\left|\Omega_{dr}\right|}
{(3|w|-1))\Omega_{m}}\right)^{\frac{1}{3|w|+1}},
\eq
as shown in Fig. \ref{fig9}. 

\begin{figure}
\includegraphics[width=\columnwidth]{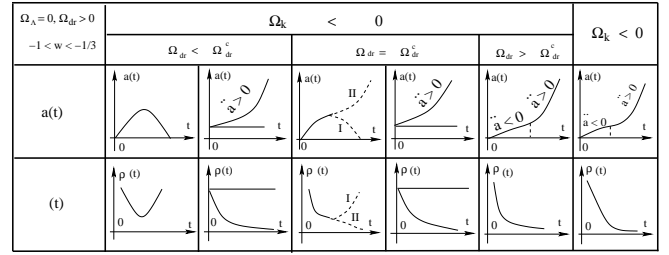}
\caption{The evolution of the universe  given by Eq.(\ref{3.22}) for $\Omega_{\Lambda} =   0, \;  \Omega_{dr} > 0 $  and $- 1 < w < - 1/3$. }
\label{fig10}
\end{figure}

When $\Omega_{k} < 0$, there exists a critical value $\Omega^{c}_{dr}$,
\bq
\lb{3.26}
\Omega^{c}_{dr} \equiv \frac{(3|w|-1)a^{2}_{max}}{3|w| + 1}\left| \Omega_{k}\right|,
\eq
so that when $\Omega_{dr}  < \Omega^{c}_{dr}$ the potential has two positive roots,
say, $a_{1}$ and $a_{2}$, where $a_{2} > a_{1}$, as shown by Curve (a) in Fig.\ref{fig9}, where
$V(a_{max},   \Omega^{c}_{dr}) = V'(a_{max},   \Omega^{c}_{dr}) = 0$. Then, we can see
that the motion can have two different kinds, depending on the choice of the initial condition
of the universe. It can start to expand from the big bang at $a(0) = 0$ until its maximal radius
$a_{1}$ and then starts to collapse. The collapsing process is exactly the time-inverse
process of the expansion, and in particular, a big crunch singularity is formed at $t = t_{s} >0$
where $a(t_{s}) = 0$. Since $dv(a)/da > 0$ for any given value of $a \in (0, a_{1}]$, we can see
that in this case the universe is decelerating, as shown in Fig.\ref{fig10}. If the universe starts
to expand at a radius $a_{i}$ where $a_{i} \ge a_{2}$, then we can see that it will expand forever,
and the corresponding acceleration is always positive. 

When $\Omega_{dr}  =  \Omega^{c}_{dr}$,
the motion can have  two different types, too, and the only difference between the last case and 
the current one is that
now the two roots $a_{1}$ and $a_{2}$ are degenerate and are all equal to $a_{max}$, as shown by Curve
(b) in Fig.\ref{fig9}. As a 
result, this point represents an unstable static point, and any kind of perturbations will lead the
universe either to collapse or to expand forever, as shown in Fig.\ref{fig10}.  When $\Omega_{dr}  
> \Omega^{c}_{dr}$, the potential is always negative, as shown by Curve
(c) in Fig.\ref{fig9}, and the universe will start to expand from
the big bang singularity at $a(0) = 0$ forever. Initially the  universe is decelerating, but once
it expands to $a_{max}$, it will be accelerating. 

When $\Omega_{k} > 0$, the potential is always negative, as shown by Curve (d) in Fig.\ref{fig9},
and the universe can start to expand from a big bang singularity at $a(0) = 0$ until $a = \infty$.
In this case there is no turning point. The universe is initially decelerating until $a = a_{max}$
and then turns to expand acceleratingly, as shown in Fig. \ref{fig10}.

 {\bf Case A.5.2) $\; \Omega_{dr} < 0$}: In this case, from Eq.(\ref{3.22}) we can see that the potential is always positive
for $\Omega_{k} < 0$, so that the motion is forbidden. When  $\Omega_{k} > 0$,
 the potential is monotonically decreasing, as shown by Curve (b) in Fig.\ref{fig5}, from which we can
 see that the potential becomes negative when $a > a_{min}$, and $\ddot{a} (a > a_{min}) > 0$,
 where $V(a_{min}) = 0$. Thus, in the present case the motion of the universe is restricted to
 $a \ge a_{min} > 0$, and no big bang or big crunch singularity is developed. The universe 
 expands from $a_{i} \ge a_{min}$ forever until $a = \infty$. Note that the spacetime is not
 singular even at  $a = \infty$, as shown in Fig.\ref{fig6}. If initially the universe is in its collapsing phase, 
 where $a_{i} > a_{min}$ and $\dot{a}(t_{i}) < 0$, a bouncing universe is also allowed.

\subsubsection{$ w = -1$}

\begin{figure}
\includegraphics[width=\columnwidth]{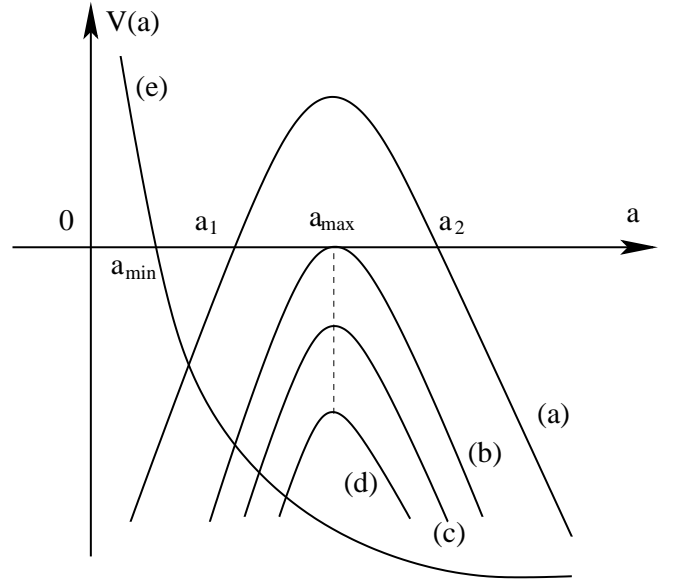}
\caption{The potential given by Eq.(\ref{3.27})   for  $\Omega_{\Lambda} =   0 $  and $w = -1$:   
(a)  $\Omega_{dr} > 0, \;  \Omega_{k} < 0,\; \Omega_{m} <  \Omega_{m}^{c}$; 
 (b)  $\Omega_{dr} > 0, \;  \Omega_{k} < 0,\; \Omega_{m} = \Omega_{m}^{c}$; 
 (c)   $\Omega_{dr} > 0, \;  \Omega_{k} < 0,\; \Omega_{m} >  \Omega_{m}^{c}$; 
(d)  $\Omega_{dr} > 0, \;  \Omega_{k} >  0$; 
and  (e)  $\Omega_{dr} < 0$.
}
\label{fig11}
\end{figure}

In this case,   Eq.(\ref{3.9}) can be written as
\bq
\lb{3.27}
V(a) = - \frac{1}{2}\left(\Omega_{k} + \Omega_{m}a^{2} + \frac{\Omega_{dr}}{a^{2}}\right),
\eq
from which we have
\bq
\lb{3.28}
V'(a) = - \frac{1}{a}\left(\Omega_{m}a^{2}  -  \frac{ \Omega_{dr}}{a^{2}}\right).
\eq
Then, it can be shown that the potential is given by Fig.\ref{fig11}, where 
Curve (a) corresponds to $\Omega_{dr} > 0, \;  \Omega_{k} < 0,\; \Omega_{m} <  \Omega_{m}^{c}$; 
Curve (b)  to $\Omega_{dr} > 0, \;  \Omega_{k} < 0,\; \Omega_{m} = \Omega_{m}^{c}$; 
Curve (c)  to $\Omega_{dr} > 0, \;  \Omega_{k} < 0,\; \Omega_{m} >  \Omega_{m}^{c}$; 
Curve (d)  to $\Omega_{dr} > 0, \;  \Omega_{k} >  0$; 
and Curve (e) to $\Omega_{dr} < 0$. 

Comparing Fig. \ref{fig11} with Fig.\ref{fig9} we can see that the  potential has the same shape in each
corresponding case of (a)-(d), so the motion of the universe in the present case can be immediately deduced
from there, and the corresponding motion of the universe in each case is given by Fig. \ref{fig10}.
In addition, comparing Curve (e) in Fig. \ref{fig11} with Curve (b) in Fig. \ref{fig5}, we can see that they
are similar, except that now the potential goes to $-\infty$ as $a \rightarrow \infty$. However, this affects
only the amplitudes of the expansion velocity and acceleration, and the main characteristics of the motion
 are the same in both cases, and is given by Fig. \ref{fig6}. However, if the universe chooses to collapse
 initially, a bouncing universe will be created, as shown by Fig.\ref{fig2a}.
 
\begin{figure}
\includegraphics[width=\columnwidth]{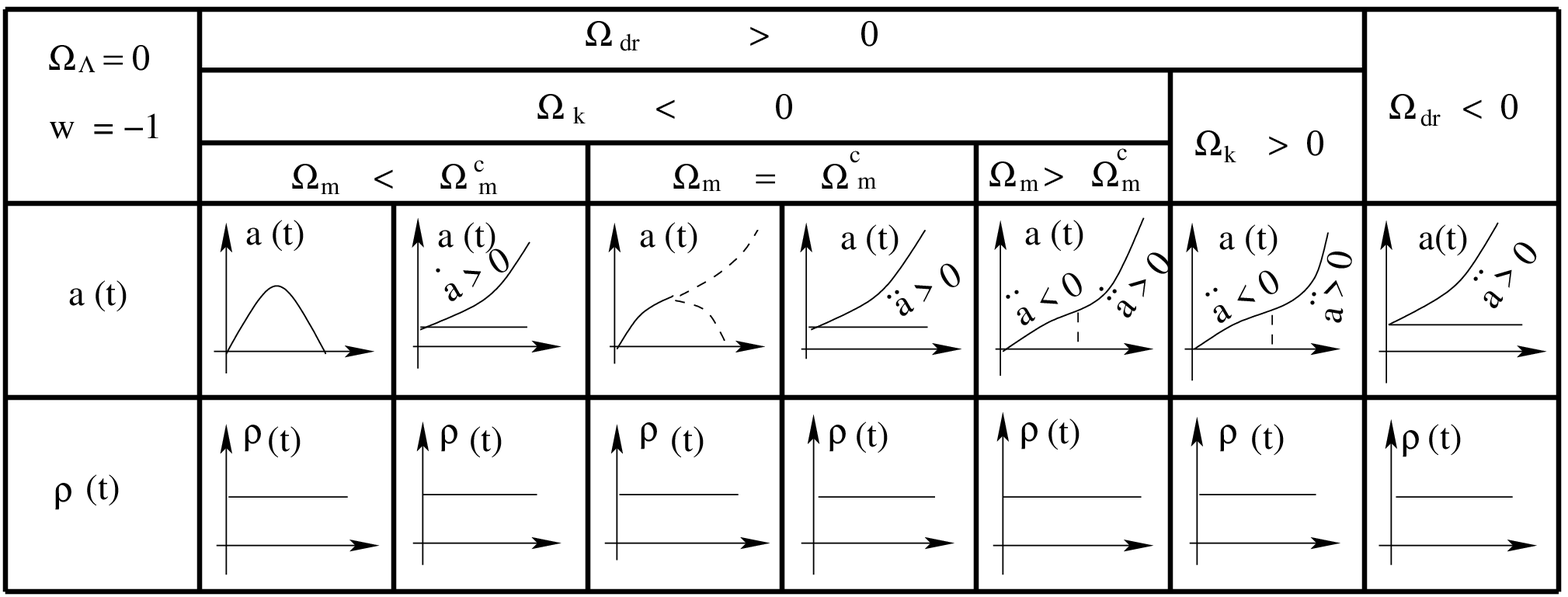}
\caption{The motion of the universe with the potential given by Eq.(\ref{3.27})  for   $\Omega_{\Lambda} =   0 $  and $w = -1$.}
\label{fig12}
\end{figure}

 In summary, the motion of the universe in this case is given by Figs. \ref{fig12} for Curves (a)-(d) in Fig.\ref{fig11},
 and by the first case in Fig.\ref{fig2a} for Curve (e).

\subsubsection{$ w < -1$}

In this case,   the corresponding fluid is usually called phantom, and Eq.(\ref{3.9}) can be written as
\bq
\lb{3.29}
V(a) = - \frac{1}{2}\left(\Omega_{k} + \Omega_{m}a^{3|w|-1} + \frac{\Omega_{dr}}{a^{2}}\right),
\eq
from which we have
\bq
\lb{3.30}
V'(a) = - \frac{1}{2}\left(\Omega_{m}(3|w| - 1)a^{3|w|-2}  -  2\frac{ \Omega_{dr}}{a^{2}}\right).
\eq
Clearly, when $\Omega_{dr} > 0$ the potential has a maximum at 
\bq
\lb{3.31}
a_{max} = \left(\frac{2\Omega_{dr}}{(3|w| -1)\Omega_{m}}\right)^{\frac{1}{3|w| + 1}}.
\eq
Meantime, if $\Omega_{k} <0$, then there exists a critical value of $\Omega_{m}^{c}$, such that
$V(a_{max}, \Omega_{m}^{c}) = V'(a_{max}, \Omega_{m}^{c}) = 0$, as shown by Curve (b) in Fig.\ref{fig11}.
Then, when $ \Omega_{m} <  \Omega_{m}^{c}$ the potential will have two positive roots, as shown by
Curve (a) in Fig.\ref{fig11}, while when  $ \Omega_{m} >  \Omega_{m}^{c}$ it is always negative, and the corresponding
curve is that of Curve (c) in Fig.\ref{fig11}. When $\Omega_{k} > 0$ the potential is always negative, and is given by
Curve (d) in Fig.\ref{fig11}.

\begin{figure}
\includegraphics[width=\columnwidth]{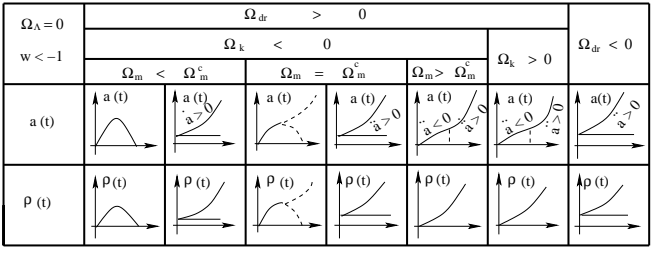}
\caption{The motion of the universe with the potential given by Eq.(\ref{3.29})  for   $\Omega_{\Lambda} =   0 $  and $w < -1$.}
\label{fig13}
\end{figure}

When $\Omega_{dr} < 0$, the potential is monotonically decreasing, and is that of Curve (e) in Fig.\ref{fig11}.
Therefore, in the present case, the motion of the universe is qualitatively the same as the corresponding one in
the last case, given by Fig. \ref{fig12} and the first case of Fig.\ref{fig2a}. The only difference is that now the 
matter part is singular as $a \rightarrow \infty$. So, now
we have a big rip singularity at  $a =  \infty$, as shown by Fig. \ref{fig13}.


\subsection{$\Omega_{\Lambda} >  0$}

When $\Omega_{\Lambda} >  0$, from Eq.(\ref{3.7}) we find that
\bqn
\lb{3.32a}
V(a) & =& - \frac{1}{2}\left(\Omega_{k}  + \Omega_{\Lambda}a^{2} +   \frac{\Omega_{dr}}{a^{2}}
+ \frac{\Omega_{m}}{a^{1+3w}}\right), \;\;\;\;\\
\lb{3.32b}
V'(a) & =& - a\left(\Omega_{\Lambda}  -   \frac{\Omega_{dr}}{a^{4}}
- \frac{(1+3w)\Omega_{m}}{2a^{3(1+w)}}\right).
\eqn
 
\subsubsection{$w > \frac{1}{3}$}

When $w > 1/3$, we first re-write Eqs.(\ref{3.32a}) and (\ref{3.32b}) in the forms,
\bqn
\lb{3.33a}
V(a) & =& - \frac{1}{2}\left(\Omega_{k}  + \Omega_{\Lambda}a^{2} \right.\nb\\
& & \left. +   \frac{1}{a^{1+3w}}\left(\Omega_{m} + \Omega_{dr}a^{3w-1}\right)\right),  \;\;\;\;\\
\lb{3.33b}
V'(a) & =& - a\left(\Omega_{\Lambda}  -   \frac{1}{a^{3(1+w)}}\right.\nb\\
& & \left.\times\left(\frac{1+3w}{2}\Omega_{m} + \Omega_{dr}a^{3w-1}\right)\right).
\eqn
To study the solution further, we consider the cases $\Omega_{dr} >0 $ and  $\Omega_{dr} < 0 $
separately.

{\bf Case B.1.1) $\; \Omega_{dr} >0$}: In this case, from Eq.(\ref{3.33a}) we find that
\bq
\lb{3.34}
V(a)= \cases{-\infty, & $a = 0$,\cr
-\infty, & $a = \infty$,\cr}
\eq
while from Eq.(\ref{3.33b}) we can see that the potential has a maximum at $a_{max}$, where
$V'(a_{max}) = 0$.  

{\bf Case B.1.1a) $\; \Omega_{dr} >0,\; \Omega_{k} < 0$}: If   $\Omega_{k} < 0$,  from 
Eq.(\ref{3.33a}) we find that we can further have $V(a_{max}) = 0$, that is, for any given $\Omega_{k} <0$
and $\Omega_{dr} > 0$, there always exists a positive solution, $\left(a_{max}, \Omega_{\Lambda}^{c}\right)$,
of the equations,
\bqn
\left|\Omega_{k}\right|  &=&    \Omega_{\Lambda}a^{2}_{max}  +   \frac{1}{a^{1+3w}_{max}}
\left(\Omega_{m} + \Omega_{dr}a^{3w-1}_{max}\right),  \;\;\;\;\nb\\
\Omega_{\Lambda}^{c}  &=&   \frac{1}{a^{3(1+w)}_{max}} 
\left(\frac{1+3w}{2}\Omega_{m} + \Omega_{dr}a^{3w-1}_{max}\right).
\eqn

When $\Omega_{\Lambda} < \Omega_{\Lambda}^{c}$, the potential will take the form of Curve (a)  in Fig. \ref{fig9},
while when  $\Omega_{\Lambda} =  \Omega_{\Lambda}^{c}$  or  $\Omega_{\Lambda} >  \Omega_{\Lambda}^{c}$,
 it will take, respectively,  the form of Curves (b) and  (c), so the  motion of the universe is similar to the corresponding 
 cases given in Fig. \ref{fig10}.
 
 {\bf Case B.1.1b) $\; \Omega_{dr} >0,\; \Omega_{k} > 0$}: In this case, it can be shown that the potential is given
 by Curve (d) in Fig. \ref{fig9}, so the motion is given by the corresponding case in Fig. \ref{fig10}.

{\bf Case B.1.2) $\; \Omega_{dr} < 0$}: In this case, it can be shown that the equations  $V(a) = 0$ and $V'(a) = 0$ 
can be cast in the form,
\bqn
\lb{3.35a}
3(1+w)\Omega_{\Lambda}a^{2} + (1+3w)\Omega_{k} =  (3w-1)\frac{\left|\Omega_{dr}\right|}{a^{2}},\\
\lb{3.35b}
 \Omega_{\Lambda}  + \frac{\left|\Omega_{dr}\right|}{a^{4}}    =  \frac{1+3w}{2}  \frac{\Omega_{m}}{a^{3(1+w)}}.
 \eqn
Clearly, for any given $\Omega_{k}$ and $\Omega_{m}$, the above equations  always have a solution $
(a, \Omega_{\Lambda}) = (a_{max} >0,
\Omega_{\Lambda}^{c} > 0)$. Then, we can see that in the current case we also have three different sub-cases
according to whether  $\Omega_{\Lambda} < \Omega_{\Lambda}^{c}$,  $\Omega_{\Lambda} =  \Omega_{\Lambda}^{c}$  
or  $\Omega_{\Lambda} >  \Omega_{\Lambda}^{c}$, for which the potential is given, respectively, by Curves (a), (b) and
(c) in Fig. \ref{fig9}, and  the  motion of the universe is given in each case by the  corresponding one given in Fig. \ref{fig10}
and Fig.\ref{fig2a}.

\subsubsection{$w = \frac{1}{3}$}

In this case, we have,
\bqn
\lb{3.36a}
V(a) & =& - \frac{1}{2}\left(\Omega_{k}  + \Omega_{\Lambda}a^{2}  +   \frac{\Omega_{\Delta}}{a^{2}}\right),\\
\lb{3.36b}
V'(a) & =& - \frac{\Omega_{\Lambda}}{a^{3}}\left(a^{4} - \frac{\Omega_{\Delta}}{\Omega_{\Lambda}}\right), 
\eqn
where $\Omega_{\Delta} \equiv  \Omega_{m} + \Omega_{dr}$.
 
{\bf Case B.2.1) $\; \Omega_{\Delta} > 0$}: In this case we find that the potential always has a maximum at
\bq
\lb{3.37}
a_{max} = \left(\frac{\Omega_{\Delta}}{\Omega_{\Lambda}}\right)^{1/4},
\eq
for which we have 
\bq
\lb{3.38}
\left. V(a) \right|_{a=a_{max}} = -\frac{1}{2}\left(\Omega_{k} + 2\sqrt{\Omega_{\Delta}\Omega_{\Lambda}}\right).
\eq
Thus, if $\Omega_{k} < 0$, a critical point $\Omega_{\Lambda}^{c}$ exists, so that $V\left(a_{max}, \Omega_{\Lambda}^{c}\right) 
= V'\left(a_{max}, \Omega_{\Lambda}^{c}\right) = 0$, where
\bq
\lb{3.39}
\Omega_{\Lambda}^{c} = \frac{\Omega_{k}^{2}}{4\Omega_{\Delta}}.
\eq
Therefore, in this case, there are three sub-cases according to whether $\Omega_{\Lambda} < \Omega_{\Lambda}^{c}$,  
$\Omega_{\Lambda} =  \Omega_{\Lambda}^{c}$  
or  $\Omega_{\Lambda} >  \Omega_{\Lambda}^{c}$, for which the potential is given, respectively, by Curves (a), (b) and
(c) in Fig. \ref{fig9}, and  the  motion of the universe is given in each case by the  corresponding one given in Figs. \ref{fig10}
and \ref{fig2a}.

When $\Omega_{k} > 0$, $V'(a)$ is always negative and the potential is given by Curve (d) in Fig. \ref{fig9}, and  the  motion 
of the universe is given  by the  corresponding case given in Fig. \ref{fig10}.

\begin{figure}
\includegraphics[width=\columnwidth]{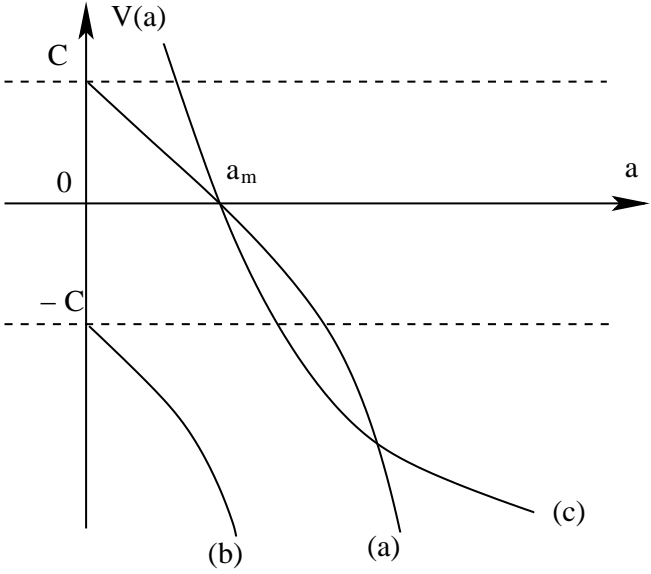}
\caption{The potential given by Eq.(\ref{3.36a}) for $w = 1/3$.  (a) $\;    \Omega_{\Lambda} >   0,\; \Omega_{\Delta} = 0, \;  \Omega_{k} < 0 $;
 (b) $\;    \Omega_{\Lambda} >   0,\; \Omega_{\Delta} = 0, \;  \Omega_{k} > 0 $; and 
  (c) $\;    \Omega_{\Lambda} >   0,\; \Omega_{\Delta} < 0$, where $C = \left|\Omega_{k}\right|/2$. }
\label{fig14}
\end{figure}

\begin{figure}
\includegraphics[width=\columnwidth]{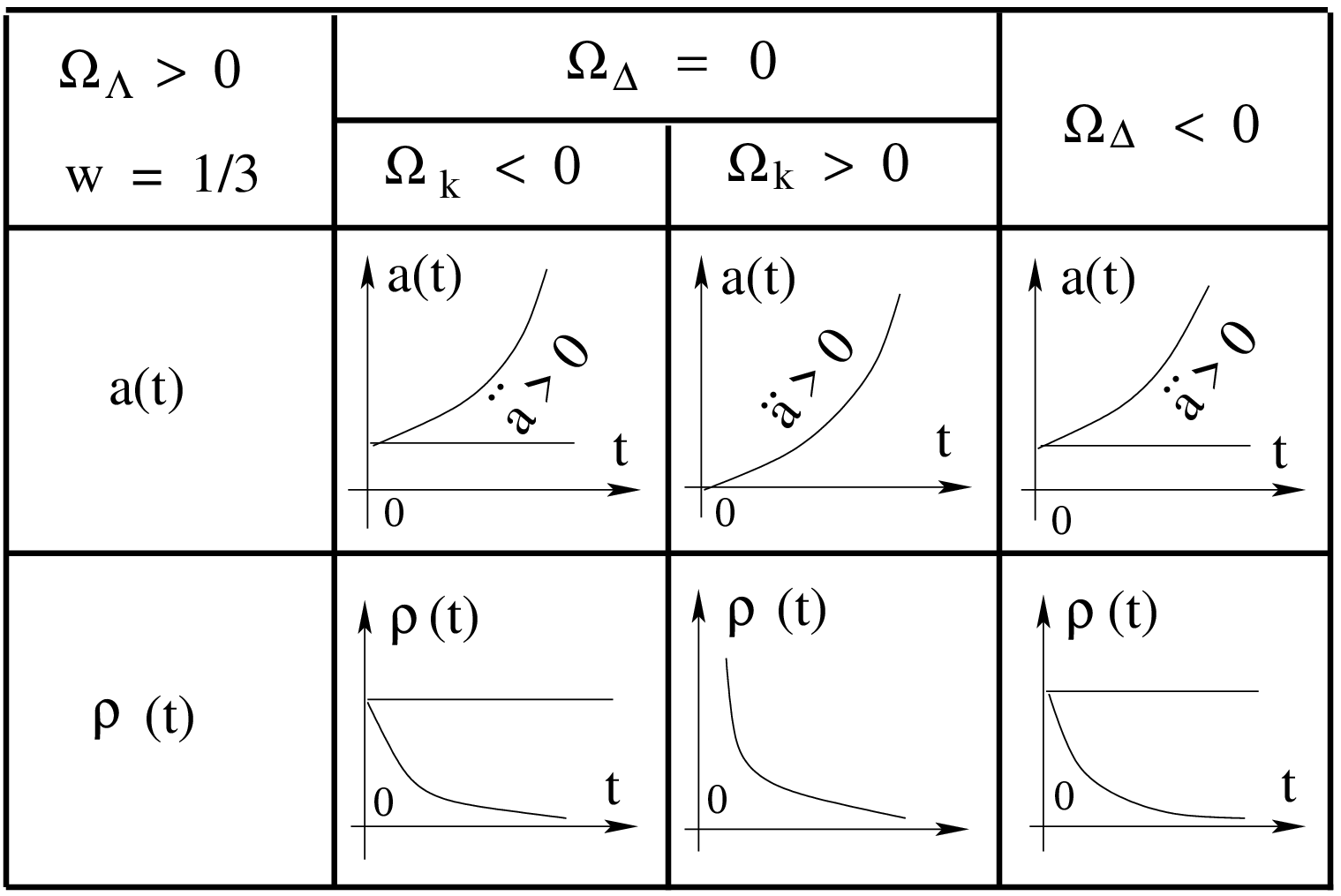}
\caption{The motion of the universe with the potential given by Eq.(\ref{3.36a})  for   $\Omega_{\Lambda} >   0 $  and $w = 1/3$.}
\label{fig15}
\end{figure}

{\bf Case B.2.2) $\; \Omega_{\Delta} = 0$}: In this case we have
\bq
\lb{3.40}
V(a) = - \frac{1}{2}\left(\Omega_{k}  + \Omega_{\Lambda}a^{2}\right), 
\eq
and the potential is given by Fig. \ref{fig14}, from which we can see that when 
$\Omega_{k} < 0$, $V(a)$ is negative only when $a > a_{m}$, where $V(a_{m}) = 0$, as shown by Curve (a) in Fig.\ref{fig14}. 
Thus, the motion in this
case is restricted to $a \ge a_{m}$. Since $dV(a)/da < 0$, we can see that now the universe is always accelerating,
as shown in Fig.\ref{fig15}. When $\Omega_{k} > 0$, the potential is always negative [as shown by Curve (b) in Fig. \ref{fig14}, 
and the universe can start to expand from the big bang at $a(0) = 0$ until $a = \infty$.

{\bf Case B.2.3) $\; \Omega_{\Delta} < 0$}: In this case we have
\bqn
\lb{3.40a}
V(a) &=& - \frac{1}{2}\left(\Omega_{k}  + \Omega_{\Lambda}a^{2}  -   \frac{\left|\Omega_{\Delta}\right|}{a^{2}}\right),\nb\\
V'(a) &=& - a \left(\Omega_{\Lambda}  +   \frac{\left|\Omega_{\Delta}\right|}{a^{4}}\right) < 0.
\eqn
Therefore, in this case the potential is a monotonically decreasing function, as shown by Curve (c) in Fig.\ref{fig14}. Then,
we can see that the motion is restricted to the region $a \ge a_{m}$. The universe   is always accelerating, and no
singularity exists in the present case, as shown in Fig. \ref{fig15}. A bouncing universe is also allowed in the present case, and
the corresponding motion is described by the first case in Fig.\ref{fig2a}.

\subsubsection{$- 1/3 < w < \frac{1}{3}$}

In this case we find that the potential and its first derivative can be written as
\bqn
\lb{3.41a}
V(a) & =& - \frac{1}{2}\left(\Omega_{k}  + \Omega_{\Lambda}a^{2} \right.\nb\\
& & \left. +   \frac{1}{a^{2}}\left( \Omega_{dr} + \Omega_{m}a^{1-3w}\right)\right),  \;\;\;\;\\
\lb{3.41b}
V'(a) & =& - a\left(\Omega_{\Lambda}  -   \frac{1}{a^{4}}\right.\nb\\
& & \left.\times\left(\Omega_{dr}  + \frac{1+3w}{2}\Omega_{m} a^{1-3w} \right)\right).
\eqn
To study this case further, we need to consider the cases $\Omega_{dr}  > 0$ and $\Omega_{dr}  <  0$ separately.

{\bf Case B.3.1) $\; \Omega_{dr}  > 0$}: In this case from the above equations we find that
\bq
\lb{3.41c}
V(a) = \cases{-\infty, & $a = 0$,\cr
-\infty, & $a = \infty$,\cr}
\eq
and that the conditions $V(a) = V'(a) = 0$ can be written as
\bqn
\lb{3.42a}
& & \frac{1+3w}{2} \Omega_{k} + \frac{3(w+1)}{2} \Omega_{\Lambda}a^{2}  =   \frac{1-3w}{2a^{2}} \Omega_{dr},\\
\lb{3.42b}
& &  \Omega_{\Lambda} = \frac{1}{a^{4}}\left(\Omega_{dr} +  \frac{1+3w}{2} \Omega_{m} a^{1-3w}\right). 
\eqn
Clearly, for any given $\Omega_{dr} > 0$ and $\Omega_{k}$, the above equation always have a positive solution
$\left(a, \Omega_{\Lambda}\right) = \left(a_{max}, \Omega_{\Lambda}^{c}\right)$, so that when $\Omega_{\Lambda}
< \Omega_{\Lambda}^{c}$, the potential is given by Curve (a) in Fig.\ref{fig9},  when $\Omega_{\Lambda}
=  \Omega_{\Lambda}^{c}$, it is given by Curve (b), and when $\Omega_{\Lambda}
> \Omega_{\Lambda}^{c}$,  it is given by Curve (c). Then, we can see that the motion in this case is given by the
corresponding case given in Fig.\ref{fig10}.

\begin{figure}
\includegraphics[width=\columnwidth]{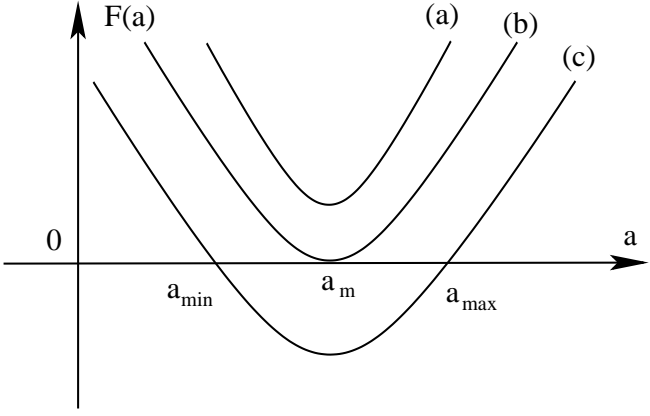}
\caption{The function $F(a)$ defined by Eq.(\ref{3.43a}) for  $\Omega_{\Lambda} > 0,\;  \Omega_{dr} < 0$ and $- 1/3 < w < 1/3$.
  (a) $\; \left|\Omega_{dr}\right| > \left|\Omega_{dr}^{c}\right|$; (b) $\; \left|\Omega_{dr}\right| = \left|\Omega_{dr}^{c}\right|$;
  and (c) $\; \left|\Omega_{dr}\right| < \left|\Omega_{dr}^{c}\right|$.}
\label{fig16a}
\end{figure}

{\bf Case B.3.2) $\; \Omega_{dr}  <  0$}: In this case from Eqs.(\ref{3.41a}) and (\ref{3.41b}) we find that
\bq
\lb{3.43}
V(a) = \cases{\infty, & $a = 0$,\cr
-\infty, & $a = \infty$,\cr}
\eq
and 
\bqn
\lb{3.43a}
V'(a) &=&-  \frac{1}{a^{3}}\left(\Omega_{\Lambda}a^{4} - \frac{1+3w}{2}\Omega_{m}a^{1-3w} + \left|\Omega_{dr}\right|\right)\nb\\
& \equiv& - \frac{1}{a^{3}}F(a).
\eqn
From the above expression we can see that there exists a critical value of $\Omega_{dr}^{c}$, so that when $\left|\Omega_{dr}\right|
> \left|\Omega_{dr}^{c}\right|$, the function $F(a)$ is always positive, and $V'(a) = 0$ has no real positive root, as shown 
by Curve (a) in Fig.\ref{fig16a}. When $\left|\Omega_{dr}\right| = \left|\Omega_{dr}^{c}\right|$, the equation $F(a) = 0$ has only
one real positive root, as shown by Curve (b),  and when $\left|\Omega_{dr}\right| < \left|\Omega_{dr}^{c}\right|$,  
$F(a) = 0$ has  two  real positive roots, as shown by Curve (c) in  Fig.\ref{fig16a}. 

On the other hand,   the conditions $V(a) = V'(a) = 0$ can be written as
\bqn
\lb{3.44a}
& & \frac{1+3w}{2} \Omega_{k} + \frac{3(w+1)}{2} \Omega_{\Lambda}a^{2}  +   \frac{1-3w}{2a^{2}}\left| \Omega_{dr}\right| 
= 0,\;\;\;\;\;\;\;\;\;\\
\lb{3.44b}
& &  \Omega_{\Lambda} +  \frac{\left| \Omega_{dr}\right|}{a^{4}}  =  \frac{(1+3w) \Omega_{m}}{2 a^{3(1+w)}}. 
\eqn
From Eq.(\ref{3.44a}) we can see that it has solution  only when  $\Omega_{k} < 0$. Thus, in the following we need to consider the
cases  $\Omega_{k} < 0$ and  $\Omega_{k} > 0$ separately.

\begin{figure}
\includegraphics[width=\columnwidth]{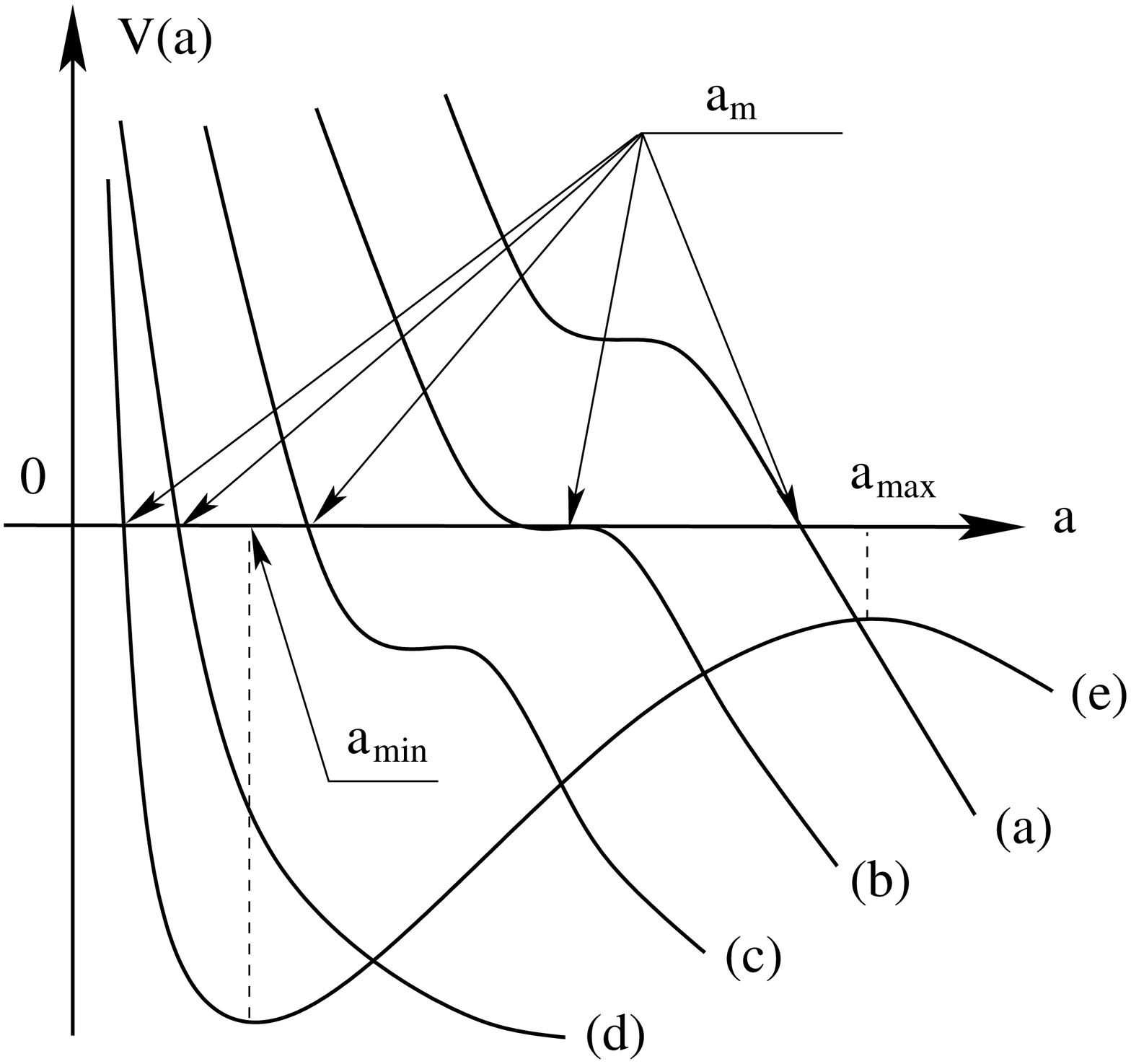}
\caption{The potential given by Eq.(\ref{3.41a}) for $- 1/3 < w < 1/3$ and $ \Omega_{dr} < 0,\;   \Omega_{k} > 0$.  (a) $\; 
  \left|\Omega_{dr}\right| = \left|\Omega_{dr}^{c}\right|,\; V(a_{min}) > 0$;  (b) $\; 
  \left|\Omega_{dr}\right| = \left|\Omega_{dr}^{c}\right|,\; V(a_{min}) = 0$; (c)  $\; 
  \left|\Omega_{dr}\right| = \left|\Omega_{dr}^{c}\right|,\; V(a_{min}) < 0$;
 (d) $\;  \left|\Omega_{dr}\right| > \left|\Omega_{dr}^{c}\right|,\; V(a_{min}) > 0$;
 and (e)  $\;  \left|\Omega_{dr}\right| < \left|\Omega_{dr}^{c}\right|$. }
\label{fig16}
\end{figure}

{\bf Case B.3.2a) $\; \Omega_{k}  >  0$}: In this case from the above we can see that 
when $\left|\Omega_{dr}\right| > \left|\Omega_{dr}^{c}\right|$,
the potential is a monotonically decreasing function, and given by Curve (d)
 in Fif.\ref{fig16}.  When $\left|\Omega_{dr}\right| = \left|\Omega_{dr}^{c}\right|$,
the potential has only one minimum, and there are three different cases, as shown by Curves (a)-(c).
 In each of these four cases (a)-(d),  we can see that the potential is negative only for $a > a_{m}$ where $a_{m}$
is the positive root of $V(a) = 0$, as shown by Fig. \ref{fig16}. Therefore, the motion of the universe is restricted to
$a \ge a_{m}$. If the universe starts to expand from $a _{i} \ge a_{m}$ it will expand forever. Since $dV(a)/da < 0$ for $a > a_{m}$, 
we can see that the universe in each of these cases is accelerating. If it is collapsing initially, then it will reach its minimal 
radius $a_{m}$ within a finite proper time. Afterwards, it will start to expand, and a bouncing universe is produced,
as described by the first case in Fig.\ref{fig2a}.  

When $\left|\Omega_{dr}\right| < \left|\Omega_{dr}^{c}\right|$, the potential has one minimum and one maximum, as shown by
Curve (e) in  Fig. \ref{fig16}. This is a very interesting case. As Cases (a)-(d), the motion of the universe is also restricted to
the region $a \ge a_{m}$. But, it is fundamentally different from these cases: The universe is accelerating
for $a \in [a_{m},\; a_{min})$ and  $a \in (a_{max},\;\infty)$, and decelerating for  $a \in (a_{min},\; a_{max})$. Therefore,
it can describe the evolution of our universe without a big bang singularity.  In particular, if the universe chooses to collapse
first at $a_{i}$, where $ a_{min} < a_{i} < a_{max}$, we can see that a scale-invariant perturbation can be produced during
this matter-dominated period \cite{wand}, and the universe will experience a bouncing once it collapses at $a_{m}$, whereby
a big bang singularity is avoided. Once it turns to expand, it will first expand acceleratingly until $a = a_{min}$. Clearly, if this
expansion is large enough, the horizon problem can be solved. Afterwards, the universe will experience a decelerating period
until $a = a_{max}$. Once this point reaches, it will expand with a positive acceleration, which may be identified with the late
cosmic acceleration.

\begin{figure}
\includegraphics[width=\columnwidth]{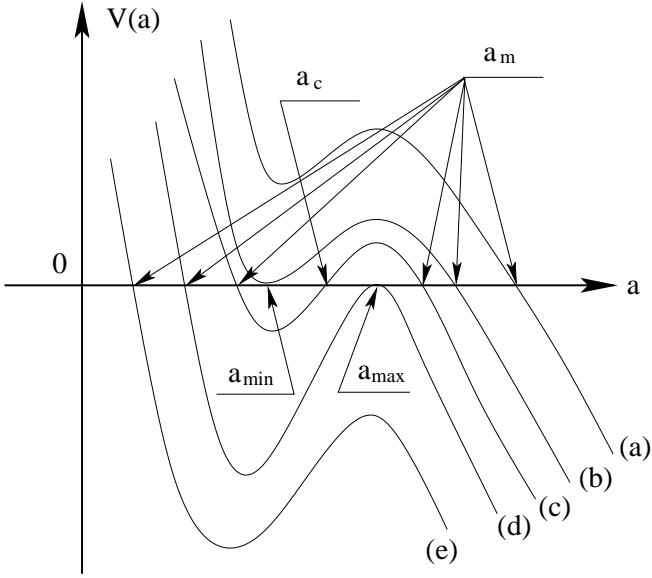}
\caption{The potential given by Eq.(\ref{3.41a}) for $ \Omega_{dr} < 0,\;   \Omega_{k} < 0,\;  \left|\Omega_{dr}\right| > \left|\Omega_{dr}^{c}\right|$, 
and  $- 1/3 < w < 1/3$.  (a) $\; V(a_{min}) > 0,\; V(a_{max}) > 0$; (b) $\; V(a_{min}) = 0,\; V(a_{max}) > 0$;
(c) $\; V(a_{min}) <  0,\; V(a_{max}) > 0$; (d) $\; V(a_{min}) <  0,\; V(a_{max}) = 0$;
and (e) $\; V(a_{min}) < 0,\; V(a_{max}) < 0$. }
\label{fig17a}
\end{figure}

{\bf Case B.3.2b) $\; \Omega_{k}  <  0$}: Now it can be shown that in the sub-cases $\left|\Omega_{dr}\right| > \left|\Omega_{dr}^{c}\right|$
and  $\left|\Omega_{dr}\right| = \left|\Omega_{dr}^{c}\right|$
the potential is quite similar to the corresponding cases given in Fig.\ref{fig16}, that is, it is given by Curve (d) for
 $\left|\Omega_{dr}\right| > \left|\Omega_{dr}^{c}\right|$ and Curves (a)-(c) for  $\left|\Omega_{dr}\right| = \left|\Omega_{dr}^{c}\right|$.
 However, when $\left|\Omega_{dr}\right| < \left|\Omega_{dr}^{c}\right|$, the conditions $V(a) = 0 = V'(a)$ have positive root for
 $a$, and now we have five different cases, as shown in Fig.\ref{fig17a}. In the case described by Curve (a) the motion of the universe
 is restricted to the region $a \ge a_{m}$, and it can represent an expanding or a bouncing universe, depending on the initial velocity
 of the universe. In the case described by Curve (b), the motion of the universe is similar to the last case, except that now a stationary
 universe also exists   at $a = a_{min}$. In the case described by Curve (c), the motion of the universe for $a \ge a_{m}$ is similar to the last 
 two cases, but now a bouncing cyclic universe exists for $a\in [a_{m},\; a_{c}]$. For the case described by Curve (d), if the universe
 starts to expand at $a_{i} > a_{m}$, it will expand forever with $\ddot{a} > 0$. If it collapse from a point $a_{i} > a_{m}$, once it reaches
 $a = a_{max}$ it will stay there. However, since it is a non-stable point, with a small perturbation, the universe either continuously collapses
 until $a = a_{m}$, or starts to expand forever. If it continuously collapse, when it reaches $a_{m}$, it will  start to expand until $a_{max}$. The 
 following motion can either continuously expand or collapse. In tis case, the universe can also move between $a_{m}$ and $a_{max}$,
 as shown in Fig.\ref{fig17a}. In the case described by Curve (e), the motion is the same as the corresponding case in Fig.\ref{fig16},
 where a bouncing cyclic universe is produced. 
  
\subsubsection{$w = - \frac{1}{3}$}

In this case we find that the potential and its first derivative can be written as
\bqn
\lb{3.45a}
V(a) & =& - \frac{1}{2}\left(\Omega_{\Delta}  + \Omega_{\Lambda}a^{2} +  \frac{\Omega_{dr}}{a^{2}}\right),  \\
\lb{3.45b}
V'(a) & =& - \frac{\Omega_{\Lambda}}{a^{3}} \left(a^{4}  -   \frac{\Omega_{dr} }{\Omega_{\Lambda}}\right).
\eqn
To study this case further, we need to consider the cases $\Omega_{dr}  > 0$ and $\Omega_{dr}  <  0$ separately.

{\bf Case B.4.1) $\; \Omega_{dr}  > 0$}: In this case from Eq.(\ref{3.45a}) we find that $V(a) \rightarrow - \infty$ as
$a \rightarrow 0$ and $a \rightarrow \infty$, while from Eq.(\ref{3.45b}) we can see that the potential has a maximum
at $a_{max} = (\Omega_{dr}/\Omega_{\Lambda})^{1/4}$, for which we have
\bq
\lb{3.46}
V\left(a_{max}\right)  = - \frac{1}{2}\left(\Omega_{\Delta}
+ 2\sqrt{\Omega_{dr}\Omega_{\Lambda}}\right).
\eq
Therefore, if $\Omega_{\Delta} < 0$, there exists a critical value of $\Omega_{\Lambda}^{c} \equiv \Omega_{\Delta}^{2}/(4\Omega_{dr}$,
for which we have $V\left(a_{max}, \Omega_{\Lambda}^{c}\right) = V'\left(a_{max}, \Omega_{\Lambda}^{c}\right) = 0$, as shown by
Curve (b) in Fig.\ref{fig9}. When $\Omega_{\Lambda} < \Omega_{\Lambda}^{c}$ the potential is given by Curve (a), and when 
 $\Omega_{\Lambda} < \Omega_{\Lambda}^{c}$it is given by Curve (c). When  $\Omega_{\Delta} >  0$, the potential is always
 negative, and described by Curve (d) in Fig.\ref{fig9}.  Thus, in these sub-cases the motion of the universe is described by 
 the corresponding cases given in Fig.\ref{fig10}. 
 
 {\bf Case B.4.2) $\; \Omega_{dr}  <  0$}:  In this case, we have
 \bq
\lb{3.47}
V(a) = \cases{\infty, & $a = 0$,\cr
-\infty, & $a = \infty$,\cr}
\eq
and $V'(a)$ is always negative. Then, the potential is given by Curve (e) in Fig.\ref{fig11}, and the corresponding motion of the 
universe is described by the corresponding case given in Fig.\ref{fig12} and the first case of Fig.\ref{fig2a}. The only difference is
 that now the matter density diverges
at the big bang singularity $a(0) = 0$.

\subsubsection{$-1 < w < -1/3$}

In this case we find that the potential and its first derivative can be written as
\bqn
\lb{3.47a}
V(a) & =& - \frac{1}{2}\left(\Omega_{k}  + \frac{\Omega_{dr}}{a^{2}} \right.\nb\\
& & \left. +   {a^{3|w|-1}}\left(\Omega_{m} + \Omega_{\Lambda}a^{3(1 - |w|)}\right)\right),  \;\;\;\;\\
\lb{3.47b}
V'(a) & =& - a\left(\Omega_{\Lambda}  -   \frac{1}{a^{4}}\right.\nb\\
& & \left.\times\left(\Omega_{dr}  -  \frac{3|w|-1}{2}\Omega_{m} a^{1-3w} \right)\right).
\eqn
To study this case further, we need to consider the cases $\Omega_{dr}  > 0$ and $\Omega_{dr}  <  0$ separately.

{\bf Case B.5.1) $\; \Omega_{dr}  > 0$}: In this case from the above equations we find that
\bq
\lb{3.48}
V(a) = \cases{-\infty, & $a = 0$,\cr
-\infty, & $a = \infty$,\cr}
\eq
and that the conditions $V(a) = V'(a) = 0$ can be written as
\bqn
\lb{3.49a}
& &  \Omega_{k} = - \frac{1}{a^{2}}\left( \Omega_{dr} + a^{3|w|+1} \right.\nb\\
& & \;\;\;\;\;\;\;\;\; \left. \times \left(\Omega_{m} + a^{3(1-|w|)} \Omega_{\Lambda}\right)\right),\\
\lb{3.49b}
& &  \Omega_{\Lambda} = \frac{1}{a^{4}}\left(\Omega_{dr} -  \frac{3|w|-1}{2} \Omega_{m} a^{3|w|+1}\right). 
\eqn
Clearly, they have solutions for positive $a$ and $\Omega_{\Lambda}$ only when $\Omega_{k} < 0$. Then, we have three different sub-cases,
as given by Curves (a), (b) and (c) in Fig. \ref{fig9}, which correspond to, respectively, $\Omega_{\Lambda} < \Omega_{\Lambda}^{c}$, 
$\Omega_{\Lambda} = \Omega_{\Lambda}^{c}$, and $\Omega_{\Lambda} > \Omega_{\Lambda}^{c}$, where $ \Omega_{\Lambda}^{c}$
is the solution of Eqs.(\ref{3.49a}) and (\ref{3.49b}). When $\Omega_{k} > 0$ the potential is always negative and is given by Curve (d)
in Fig. \ref{fig9}, so the motion of the universe is described by the corresponding cases in Fig.\ref{fig10}.

{\bf Case B.5.2) $\; \Omega_{dr}  < 0$}: In this case  
we find that
\bq
\lb{3.50}
V(a) = \cases{\infty, & $a = 0$,\cr
-\infty, & $a = \infty$,\cr}
\eq
and $V'(a) < 0$. Therefore, now the potential is monotonically decreasing, and is given by Curve (e) in Fig.\ref{fig11}. Then, the motion of the 
universe  in this case is given by the last case described in Fig.\ref{fig12}, except that now we still have $\rho(0) = \infty$ and 
 $\rho(\infty) = 0.$ Certainly, a bouncing universe is also allowed. 
 
 \subsubsection{$w = -1$}
 
 In this case, we find that
\bqn
\lb{3.51a}
V(a) & =& - \frac{1}{2}\left(\Omega_{k} + \Omega_{m, \Lambda} a^{2}  + \frac{\Omega_{dr}}{a^{2}}\right),  \;\;\;\;\\
\lb{3.51b}
V'(a) & =& - a\left(\Omega_{m, \Lambda}   -  \frac{\Omega_{dr}}{ a^{4}}\right),
\eqn
where $\Omega_{m, \Lambda} \equiv \Omega_{m} + \Omega_{\Lambda} > 0$.
Comparing the above equations with Eqs.(\ref{3.27}) and (\ref{3.28}) in the case $\Omega_{\Lambda} = 0$, we can see
that by exchanging $\Omega_{m}$ with $\Omega_{m, \Lambda}$, we can get one set of the equations from the other. Therefore, the 
motion of the universe in the present case can be immediately deduced from the corresponding ones given in Figs. \ref{fig11},
\ref{fig12}, and \ref{fig2a}.

\subsubsection{$w < -1$}

In this case we find that the potential and its first derivative can be written as
\bqn
\lb{3.52a}
V(a) & =& - \frac{1}{2}\left(\Omega_{k}  + \frac{\Omega_{dr}}{a^{2}} \right.\nb\\
& & \left. +   a^{2}\left(\Omega_{\Lambda} + \Omega_{m}a^{3(|w|-1)}\right)\right),  \;\;\;\;\\
\lb{3.52b}
V'(a) & =& - a\left(\Omega_{\Lambda}  -   \frac{\Omega_{dr}}{a^{4}}\right.\nb\\
& &  \left. +  \frac{3|w|-1}{2}\Omega_{m} a^{3|w|-1}\right).
\eqn
 
{\bf Case B.7.1) $\; \Omega_{dr}  > 0$}: In this case  we find that
\bq
\lb{3.53}
V(a) = \cases{-\infty, & $a = 0$,\cr
-\infty, & $a = \infty$,\cr}
\eq
and that the conditions $V(a) = V'(a) = 0$ can be written as
\bqn
\lb{3.54a}
& &  \Omega_{k} = -  a^{2}\left(\Omega_{\Lambda} + a^{3|w|-1} \Omega_{m}\right)  
- \frac{\Omega_{dr}}{a^{2}},\\
\lb{3.54b}
& &  \Omega_{\Lambda}  +  \frac{\Omega_{dr}}{a^{4}} +    \frac{3|w| -1}{2} \Omega_{m} a^{3|w|-1}. 
\eqn
Clearly, they have solutions for positive $a$ and $\Omega_{\Lambda}$ only when $\Omega_{k} < 0$. 
When $\Omega_{k} > 0$, the potential is strictly negative, and is given by Curve (d) in Fig.\ref{fig11}.
When $\Omega_{k} < 0$, Eqs.(\ref{3.54a}) and (\ref{3.54b}) have a unique solution $\left(a_{max}, \Omega_{\Lambda}^{c}\right)$,
so that the potential is given, respectively, by  Curves (a), (b) and (c) in Fig. \ref{fig11}, which correspond to $\Omega_{\Lambda}
 < \Omega_{\Lambda}^{c}$,  $\Omega_{\Lambda} = \Omega_{\Lambda}^{c}$, and $\Omega_{\Lambda} > \Omega_{\Lambda}^{c}$.
Therefore, the motion of the universe can be deduced from there and can be shown that it is given by Fig.\ref{fig13}.
 
{\bf Case B.7.2) $\; \Omega_{dr}  < 0$}: In this case  
we find that
\bq
\lb{3.55}
V(a) = \cases{\infty, & $a = 0$,\cr
-\infty, & $a = \infty$,\cr}
\eq
and $V'(a) < 0$. Therefore, now the potential is monotonically decreasing, and is given by Curve (e) in Fig.\ref{fig11}. Then, the motion of the 
universe  in this case is given by the last case described in Fig.\ref{fig13} and the first case in Fig.\ref{fig2a}. The only difference is
that now the spacetime is singular at $a = \infty$.

\subsection{$\Omega_{\Lambda} < 0$}

When $\Omega_{\Lambda} <  0$, from Eq.(\ref{3.7}) we find that
\bqn
\lb{3.56a}
V(a) & =& - \frac{1}{2}\left(\Omega_{k}  - \left|\Omega_{\Lambda}\right|a^{2} +   \frac{\Omega_{dr}}{a^{2}}
+ \frac{\Omega_{m}}{a^{1+3w}}\right), \;\;\;\;\\
\lb{3.56b}
V'(a) & =&  a\left(\left|\Omega_{\Lambda}\right|  +  \frac{1}{a^{3(1+w)}}\right.\nb\\
& & \left. \times\left(\frac{1+3w}{2}\Omega_{m} + a^{3w-1}\Omega_{dr}\right)\right). 
\eqn
 
\subsubsection{$w > \frac{1}{3}$}

In this case, we have  
\bq
\lb{3.57}
V(a) = \cases{-\infty, & $a = 0$,\cr
\infty, & $a = \infty$.\cr}
\eq

\begin{figure}
\includegraphics[width=\columnwidth]{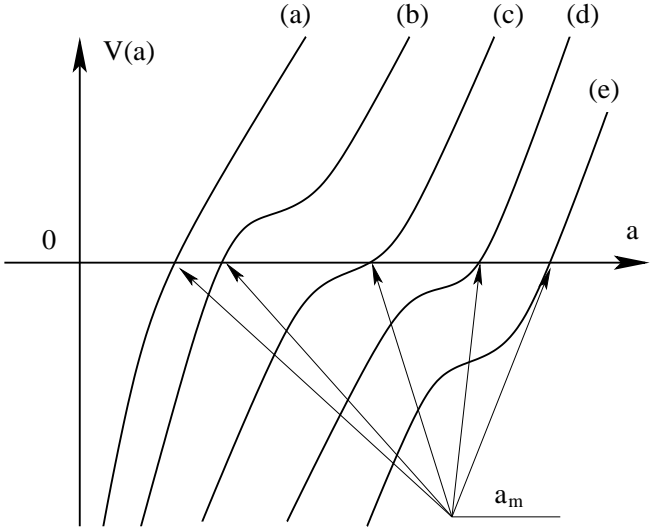}
\caption{The potential given by Eq.(\ref{3.56a}) for $  w >1/3$.  (a) $\;   \Omega_{dr} > 0$; 
 (b) $\;   \Omega_{dr} < 0,\; \Omega_{k} > 0,\;   \left| \Omega_{\Lambda}\right|  >     \left| \Omega_{\Lambda}^{c}\right|$; 
 (c) $\;   \Omega_{dr} < 0,\; \Omega_{k} > 0,\;   \left| \Omega_{\Lambda}\right|  =     \left| \Omega_{\Lambda}^{c}\right|$; 
 (d) $\;   \Omega_{dr} < 0,\; \Omega_{k} > 0,\;   \left| \Omega_{\Lambda}\right|  <     \left| \Omega_{\Lambda}^{c}\right|$; 
 and
  (e) $\;   \Omega_{dr} < 0,\; \Omega_{k} < 0$.   }
\label{fig17}
\end{figure}

When $\Omega_{dr} > 0$, from Eq.(\ref{3.56b}) we can see that $dV(a)/da$ is always positive, and the potential is monotonically increasing, as
shown by Curve (a) in Fig. \ref{fig17}. Thus, the motion now is restricted to $a \le a_{m}$.

When $\Omega_{dr} < 0$, the equations $V(a) = 0$ and $ V'(a) = 0 $  can be written in the forms,
\bqn
\lb{3.58a}
& & \Omega_{k}  =  \frac{3(1+w)\left|\Omega_{\Lambda}\right|}{1+ 3w} a^{2} + \frac{(3w-1)\left|\Omega_{dr}\right|}{(1+ 3w) a^{2}},\\
\lb{3.58b}
& & \left|\Omega_{\Lambda}\right| + \frac{(1+3w)\Omega_{m}}{2 a^{3(1+w)}} = \frac{\left|\Omega_{dr}\right|}{a^{4}},
\eqn
which have solutions for $a > 0$ and $\Omega_{\Lambda} < 0$ only when $\Omega_{k} > 0$. In the latter case, there exist a critical value
$\left|\Omega_{\Lambda}^{c}\right|$, for which, when $\left|\Omega_{\Lambda}\right| > \left|\Omega_{\Lambda}^{c}\right|$, the potential is given by
Curve (b), when $\left|\Omega_{\Lambda}\right| = \left|\Omega_{\Lambda}^{c}\right|$,  by
Curve (c), and  when $\left|\Omega_{\Lambda}\right| < \left|\Omega_{\Lambda}^{c}\right|$,  by
Curve (d), in Fig.\ref{fig17}. When $\Omega_{k} < 0$, Eqs.(\ref{3.58a}) and (\ref{3.58b}) have not positive solutions for $a$ and 
$\left|\Omega_{\Lambda}^{c}\right|$, and the potential is given by Curve (e) in Fig.\ref{fig17}.

In all the above cases, we can see that the motion is always restricted to $a \le a_{m}$ where $a_{m}$ is the unique positive solution of $V(a) = 0$,
as shown in Fig.\ref{fig17}. Therefore, now the universe starts to expand from the big bang at $a(0) = 0$ to its maximal radius $a_{m}$, and then 
starts to collapse until a big crunch singularity is formed at the moment $t_{s}$ where $a(t_{s}) = 0$. Note that for the potential given by Curves (a)
and (b), the universe is always decelerating. For the one given by Curve (c), the point $a= a_{m}$ is a stationary point, as now we have $\dot{a}
= 0 = \ddot{a}$ at this point. However, as shown there, this point is not stable, and with a small perturbation, the universe will collapse towards 
$a = 0$ and finally a big crunch singularity is formed there. For the potential given by Curves (d) and (e), there exist a point $a = a_{min} < a_{m}$,
at which $\ddot{a}  = - H^{2}_{0}dV(a)/da = 0$, as shown by these curves. However, the universe will immediately turns to decelerate.

\subsubsection{$w = \frac{1}{3}$}

In this case, we have  
\bqn
\lb{3.60a}
V(a) & =& - \frac{1}{2}\left(\Omega_{k}  - \left|\Omega_{\Lambda}\right|a^{2} +   \frac{\Omega_{\delta}}{a^{2}}\right), \;\;\;\;\\
\lb{3.60b}
V'(a) & =&  a\left(\left|\Omega_{\Lambda}\right|   + \frac{\Omega_{\delta}}{a^{4}}\right). 
\eqn
 
\begin{figure}
\includegraphics[width=\columnwidth]{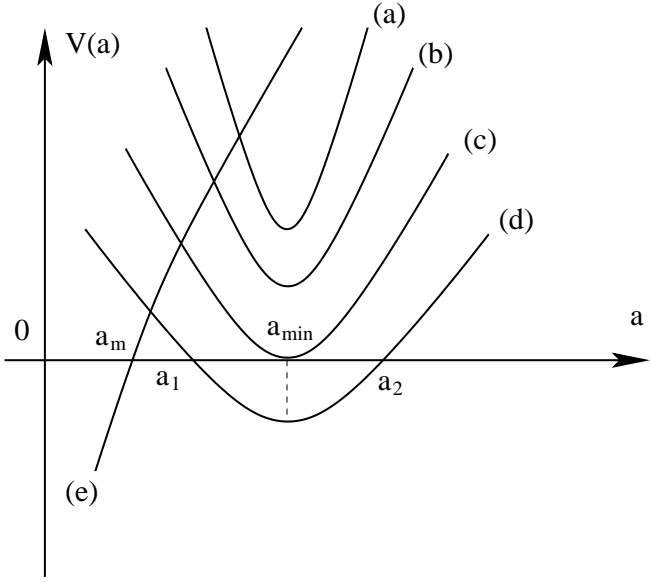}
\caption{The potential given by Eq.(\ref{3.60a}) for $  w =1/3$.  (a) $\;   \Omega_{k} < 0, \; \Omega_{\delta} < 0$; 
 (b) $\; \Omega_{k} > 0, \; \Omega_{\delta} < 0,\;  \left| \Omega_{\Lambda}\right|  >     \left| \Omega_{\Lambda}^{c}\right|$; 
 (c) $\;    \Omega_{k} > 0, \; \Omega_{\delta} < 0,\;   \left| \Omega_{\Lambda}\right|  =     \left| \Omega_{\Lambda}^{c}\right|$; 
 (d) $\;   \Omega_{k} > 0, \; \Omega_{\delta} < 0,\;   \left| \Omega_{\Lambda}\right|  <     \left| \Omega_{\Lambda}^{c}\right|$;
 and    (e) $\;    \Omega_{\delta} > 0$. }
\label{fig18}
\end{figure}

{\bf Case C.2.1) $\; \Omega_{\delta} < 0$}: In this case, we have 
\bq
\lb{3.61}
V(a) = \cases{\infty, & $a = 0$,\cr
\infty, & $a = \infty$.\cr}
\eq
When $\Omega_{k} < 0$, the potential is always positive, as shown by Curve (a) Fig. \ref{fig18}, and the motion is forbidden.

When $\Omega_{k} > 0$, $V(a) = 0 = V'(a) $ have the solution,
\bqn
\lb{3.62}
a_{min} &=&\left|\frac{2\Omega_{\delta}}{\Omega_{k}}\right|^{1/2},\nb\\
\left|\Omega_{\Lambda}^{c}\right| &=& \frac{\Omega_{k}^{2}}{4\left|\Omega_{\delta}\right|}.
\eqn
If $\left|\Omega_{\Lambda}\right| > \left|\Omega_{\Lambda}^{c}\right|$, the potential is positive, as shown by Curve (b) in 
Fig.\ref{fig18}, and the motion is forbidden. If $\left|\Omega_{\Lambda}\right| = \left|\Omega_{\Lambda}^{c}\right|$, the potential 
is also positive, except for the point $a = a_{min}$, at which we have $V(a_{min}) = 0$, as shown by Curve (c). At this
point we have $\dot{a} = 0 = \ddot{a}$, and it represents a stable stationary point,  similar to the case
described by Curve (b) in Fig.\ref{fig7}. If $\left|\Omega_{\Lambda}\right| <  \left|\Omega_{\Lambda}^{c}\right|$, the potential 
is negative only for the range $a \in (a_{1}, a_{2})$, as shown by Curve (d) in Fig.\ref{fig18}, which is also similar to Curve
(c) in Fig.\ref{fig7}, and the motion of the universe is described by the corresponding case in Fig.\ref{fig8}.

{\bf Case C.2.2) $\; \Omega_{\delta} > 0$}: In this case, we have 
\bq
\lb{3.61b}
V(a) = -\cases{\infty, & $a = 0$,\cr
\infty, & $a = \infty$,\cr}
\eq
and $V'(a) >0$, that is, now the potential is a monotonically increasing function, as shown by Curve (e) in Fig.\ref{fig18}, which is 
similar to Curve (a) in Fig.\ref{fig17}. Therefore, the motion of the universe in this case is similar to that case.

\subsubsection{$ - 1/3 < w < 1/3$}

In this case we find that

\bqn
\lb{3.62a}
V(a) & =& - \frac{1}{2}\left(\Omega_{k}  - \left|\Omega_{\Lambda}\right|a^{2} \right.\nb\\
&& \left.  +   \frac{1}{a^{2}} \left(\Omega_{dr} +  \Omega_{m}a^{1-3w}\right)\right), \;\;\;\;\\
\lb{3.62b}
V'(a) & =&  a\left(\left|\Omega_{\Lambda}\right|  +  \frac{1}{a^{4}}\right.\nb\\
& & \left. \times\left( \Omega_{dr} + \frac{1+3w}{2}\Omega_{m}a^{1-3w}\right)\right). 
\eqn

{\bf Case C.3.1) $\; \Omega_{\delta} > 0$}: In this case, we have 
\bq
\lb{3.63}
V(a) = -\cases{\infty, & $a = 0$,\cr
\infty, & $a = \infty$,\cr}
\eq
and $V'(a) >0$, so that    the potential now is a monotonically increasing function, as shown by Curve (e) in Fig.\ref{fig18}. 
Therefore, the motion of the universe in this case is similar to that one.

{\bf Case C.3.2) $\; \Omega_{\delta} < 0$}: In this case, we have 
\bq
\lb{3.63b}
V(a) = \cases{\infty, & $a = 0$,\cr
\infty, & $a = \infty$,\cr}
\eq
and $V(a) = 0 = V'(a)$ can be written as
\bqn
\lb{3.64}
&&  \Omega_{k}  + \frac{(1-3w) \left|\Omega_{dr}\right|}{(1+3w)a^{2}} = \frac{3(1+ w) \left|\Omega_{\Lambda}\right|}{(1+3w)}a^{2},\\
&&  \left|\Omega_{\Lambda}\right|    + \frac{(1+3w) \Omega_{m}}{2a^{1+3w}} = \frac{\left|\Omega_{dr}\right|}{ a^{4}}.
\eqn
Clearly, for any given $\Omega_{k}$ and $\Omega_{m}$, the above equations always have a positive solution $(a_{min},
 \Omega_{\Lambda}^{c})$, so that when $\left|\Omega_{\Lambda}\right| > \left|\Omega_{\Lambda}^{c}\right|$, the potential 
is always positive, and the motion is forbidden, as shown by Curve (b) in Fig.\ref{fig18}. When $\left|\Omega_{\Lambda}\right| = 
\left|\Omega_{\Lambda}^{c}\right|$, the only possible motion is $a = a_{min}$, at which we have $\dot{a} = 0 = \ddot{a}$, 
as shown by Curve (c). When  $\left|\Omega_{\Lambda}\right| <  \left|\Omega_{\Lambda}^{c}\right|$, the potential 
is negative in the range $a \in (a_{1}, a_{2})$, as shown by Curve (d) in Fig.\ref{fig18}. Therefore, the motion of the universe in the
current case can be deduced from the corresponding ones given in Fig.\ref{fig18}.

\subsubsection{$  w = - 1/3$}

In this case, we find that
\bqn
\lb{3.65a}
V(a) & =& - \frac{1}{2}\left(\Omega_{\Delta}  - \left|\Omega_{\Lambda}\right|a^{2}  
     +  \frac{\Omega_{dr}}{a^{2}}\right), \;\;\;\;\\
\lb{3.65b}
V'(a) & =&  \frac{\left|\Omega_{\Lambda}\right|}{a^{3}} \left(a^{4}  
 +  \frac{ \Omega_{dr}}{\left|\Omega_{\Lambda}\right|}\right).
\eqn
Thus, when $\Omega_{dr} > 0$ we have $V'(a) > 0$ and 
\bq
\lb{3.66}
V(a) = \cases{-\infty, & $a = 0$,\cr
\infty, & $a = \infty$,\cr}
\eq
that is, the potential now is a monotonically increasing function, given by Curve (e) in Fig.\ref{fig18}, and the motion of the universe
is restricted to $a \le a_{m}$. The universe starts to expand at the big bang $a(0) = 0$ until its maximal radius $a_{m}$, and then 
starts to collapse until a big crunch singularity is formed at the moment $t_{s}$ where $t_{s}$ is given by $a(t_{s}) = 0$.

When $\Omega_{dr} < 0$, we have 
\bq
\lb{3.67}
V(a) = \cases{\infty, & $a = 0$,\cr
\infty, & $a = \infty$,\cr}
\eq
and the potential has a minimum at
\bq
\lb{3.68}
a_{min} = \left|\frac{\Omega_{dr}}{\Omega_{\Lambda}}\right|^{1/4},
\eq
at which we have
\bq
\lb{3.69}
V\left(a_{min}\right) = -\frac{1}{2}\left(\Omega_{\Delta} - 2 \left|\Omega_{dr}\Omega_{\Lambda}\right|\right).
\eq
Clearly, if $\Omega_{\Delta} \le 0$, the potential is always positive, and the motion is forbidden. When $\Omega_{\Delta} > 0$,
there exist a critical value $ \Omega_{\Lambda}^{c}$, such that the potential is always positive for $\left|\Omega_{\Lambda}\right| 
> \left|\Omega_{\Lambda}^{c}\right|$,  and the motion is forbidden, as shown by Curve (b) in Fig.\ref{fig18}. 
When $\left|\Omega_{\Lambda}\right| = \left|\Omega_{\Lambda}^{c}\right|$, the only possibility of the motion is  stationary point
$a = a_{min}$, as shown by Curve (c)  in Fig.\ref{fig18}. When  $\left|\Omega_{\Lambda}\right| <  \left|\Omega_{\Lambda}^{c}\right|$, 
the potential is negative in the range $a \in (a_{1}, a_{2})$, as shown by Curve (d) in Fig.\ref{fig18}.

\subsubsection{$  -1 < w < - 1/3$}

In this case the potential and its derivative can be written as
\bqn
\lb{3.70a}
V(a) & =& - \frac{1}{2}\left(\Omega_{k}  +    \frac{\Omega_{dr}}{a^{2}} + a^{3|w|-1} \right.\nb\\
& & \left. \times \left(\Omega_{m} -   \left|\Omega_{\Lambda}\right|a^{3(1 - |w|)}\right)\right),\\
\lb{3.70b}
V'(a) & =&  a\left(\left|\Omega_{\Lambda}\right|  +  \frac{ \Omega_{dr}}{a^{4}}
- \frac{(3|w|-1)\Omega_{m}}{2a^{3(1-|w|)}}\right). 
\eqn

{\bf Case C.5.1) $\; \Omega_{dr} > 0$}: In this case, we have 
\bq
\lb{3.71}
V(a) = \cases{- \infty, & $a = 0$,\cr
\infty, & $a = \infty$,\cr}
\eq
and $V(a) = V'(a) = 0$ yield
\bqn
\lb{3.72a}
&& \Omega_{k} + \Omega_{m}a^{3|w| -1} + \frac{\Omega_{dr}}{a^{2}}  =  \left|\Omega_{\Lambda}\right| a^{2},\\
\lb{3.72b}
&& \left|\Omega_{\Lambda}\right|  +  \frac{ \Omega_{dr}}{a^{4}} = \frac{(3|w|-1)\Omega_{m}}{2a^{3(1-|w|)}}.
\eqn
Clearly, for any given $ \Omega_{k}$ and $\Omega_{m}$, there always exists a solution $(a,  \left|\Omega_{\Lambda}\right|)
= (a_{min} > 0,  \left|\Omega_{\Lambda}^{c}\right|)$, so that when   $\left|\Omega_{\Lambda}\right| 
> \left|\Omega_{\Lambda}^{c}\right|$,  the potential is given by Curve (b) in Fig.\ref{fig17}, 
when   $\left|\Omega_{\Lambda}\right|  =  \left|\Omega_{\Lambda}^{c}\right|$, it is given by Curve (c),
and when   $\left|\Omega_{\Lambda}\right|  <  \left|\Omega_{\Lambda}^{c}\right|$, it is given by Curve (d).
Then, the motion of the universe in this case can be deduced from the corresponding cases given in Fig.\ref{fig17}.

{\bf Case C.5.1) $\; \Omega_{dr} < 0$}: In this case, we have
\bq
\lb{3.73}
V(a) = \cases{ \infty, & $a = 0$,\cr
\infty, & $a = \infty$,\cr}
\eq
and $V(a) = V'(a) = 0$ yield the same equations (\ref{3.72a}) and (\ref{3.72b}) but now with $ \Omega_{dr} < 0$, 
which show that  for any given $ \Omega_{k}, \; \Omega_{dr} < 0$ and $\Omega_{m}$, they also have a unique solution,
$(a,  \left|\Omega_{\Lambda}\right|) = (a_{min} > 0,  \left|\Omega_{\Lambda}^{c}\right|)$,but now when   $\left|\Omega_{\Lambda}\right| 
> \left|\Omega_{\Lambda}^{c}\right|$,  the potential is always positive, as shown by given by Curve (b) in Fig.\ref{fig18}.
When   $\left|\Omega_{\Lambda}\right|  =  \left|\Omega_{\Lambda}^{c}\right|$, it is given by Curve (c),
and when   $\left|\Omega_{\Lambda}\right|  <  \left|\Omega_{\Lambda}^{c}\right|$, it is given by Curve (d)  in Fig.\ref{fig18}.
Then, the motion of the universe in this case can be deduced from the corresponding cases given there.

\subsubsection{$  w = -1$}

In this case we have
\bqn
\lb{3.74a}
V(a) & =& - \frac{1}{2}\left(\Omega_{k}  +  \Omega_{m,\Lambda}a^{2} +    \frac{\Omega_{dr}}{a^{2}}\right),\\
\lb{3.74b}
V'(a) & =& -  \frac{\Omega_{m,\Lambda}}{a^{3}}\left(a^{4} -  \frac{ \Omega_{dr}}{\Omega_{m,\Lambda}}\right),
\eqn
where $\Omega_{m,\Lambda}$ is defined in Eqs.(\ref{3.51a}) and (\ref{3.51b}), but now can be positive, zero
or negative. We also have the possibilities, $\Omega_{dr} > 0$ and $\Omega_{dr} < 0$. In the following we consider
each of these cases  separately.

{\bf Case C.6.1) $\;\Omega_{m,\Lambda} > 0, \; \Omega_{dr} > 0$}:  In this case, from Eq.(\ref{3.74a}) we find that
that
\bq
\lb{3.75}
V(a) = \cases{- \infty, & $a = 0$,\cr
-\infty, & $a = \infty$,\cr}
\eq
and $V'(a ) = 0$ has the solution, $a_{min} = (\Omega_{dr}/\Omega_{m,\Lambda})^{1/4}$, at which 
\bq
\lb{3.76}
V\left(a_{max}\right)  =-  \frac{1}{2}\left(\Omega_{k}  +  2\left(\left|\Omega_{m,\Lambda} \Omega_{dr} \right|\right)^{1/2}\right).
\eq
Thus, if $\Omega_{k} > 0$, we have $V\left(a_{max}\right) < 0$, and the potential is strictly negative, as shown by Curve (d)
in  Fig.\ref{fig11}. If $\Omega_{k} <  0$, we have $V\left(a_{max}, \Omega^{c}_{m,\Lambda}\right) = 0$, where
$\Omega^{c}_{m,\Lambda} = \Omega^{2}_{k}/(4\Omega_{dr}$, as shown by Curve (b) in Fig.\ref{fig11}. When 
$\Omega_{m,\Lambda} > \Omega^{c}_{m,\Lambda}$, the potential is given by Curve (c) there, for which we have $V(a) < 0$
for any given $a$. When $\Omega_{m,\Lambda} < \Omega^{c}_{m,\Lambda}$, the potential is given by Curve (a)  in Fig.\ref{fig11}, 
from which we can see that $V(a) < 0$ is only possible when $a \in (a_{1}, a_{2})$ where $a_{1,2}$ are two positive roots of
$V(a) = 0$, as shown there. Therefore, in this case the motion of the universe can be deduced from Fig.\ref{fig11}.

{\bf Case C.6.2) $\;\Omega_{m,\Lambda} > 0, \; \Omega_{dr} < 0$}:  In this case, from Eqs.(\ref{3.74a}) and (\ref{3.74b}) we find that
that
\bq
\lb{3.77}
V(a) = \cases{+ \infty, & $a = 0$,\cr
-\infty, & $a = \infty$,\cr}
\eq
and $V'(a ) < 0$, that is, now the potential is  monotonically decreasing, and the motion of the universe is possible only in the 
range $a \ge a_{min}$, as shown by Curve (e) in Fig.\ref{fig11}.

{\bf Case C.6.3) $\;\Omega_{m,\Lambda} = 0, \; \Omega_{dr} > 0$}:  In this case, from Eqs.(\ref{3.74a}) and (\ref{3.74b}) we find that
that
\bq
\lb{3.78}
V(a) = \cases{- \infty, & $a = 0$,\cr
-\Omega_{k}/2, & $a = \infty$,\cr}
\eq
and $V'(a ) > 0$. Thus, now the   potential becomes  monotonically increasing. When $\Omega_{k} < 0$, the potential is given by Curve
(a) in Fig.\ref{fig3}, from which we can see that the motion of the universe now is restricted to $a \le a_{m}$, where $V(a_{m}) = 0$.
When $\Omega_{k} > 0$, the potential is given by Curve (b) in Fig.\ref{fig3}, in which the potential is always negative, and the universe
starts to expand from a big bang until it reaches $a = \infty$ with an infinite proper time. The motion is always decelerating as now
$\ddot{a} = - H^{2}_{0}dV(a)/da < 0$.

{\bf Case C.6.4) $\;\Omega_{m,\Lambda} = 0, \; \Omega_{dr} < 0$}:  In this case, from Eqs.(\ref{3.74a}) and (\ref{3.74b}) we find that
that
\bq
\lb{3.79}
V(a) = \cases{ \infty, & $a = 0$,\cr
-\Omega_{k}/2, & $a = \infty$,\cr}
\eq
and $V'(a ) < 0$. Then, it can be shown that  when $\Omega_{k} < 0$, the potential is always positive, and given by Curve
(a) in Fig.\ref{fig5}. Therefore, in this case the motion is  forbidden.  When $\Omega_{k} > 0$, the potential is given by Curve 
(b) in Fig.\ref{fig5}, in which the potential is negative only when $a > a_{min}$, the universe is always accelerating, starting from 
a non-singular point $a_{i} \ge a_{min}$. The universe is also free from singularity at $a = \infty$. 

{\bf Case C.6.5) $\;\Omega_{m,\Lambda} <  0, \; \Omega_{dr} > 0$}:  In this case,   we find that
that
\bq
\lb{3.80}
V(a) = \cases{ - \infty, & $a = 0$,\cr
\Omega_{k}/2, & $a = \infty$,\cr}
\eq
and $V'(a ) > 0$. It can be shown that  for any given  $\Omega_{k}$, the potential is  given by Curve
(a) in Fig.\ref{fig17}. Then, the motion of the universe can be deduced from there.

{\bf Case C.6.6) $\;\Omega_{m,\Lambda} <  0, \; \Omega_{dr} < 0$}:  In this case,   we find that
that
\bq
\lb{3.81}
V(a) = \cases{  \infty, & $a = 0$,\cr
\Omega_{k}/2, & $a = \infty$,\cr}
\eq
and $V'(a ) = 0$ has the solution $a_{min} = \left|\Omega_{dr}/\Omega_{m,\Lambda}\right|^{1/4}$, at which we have
\bq
\lb{3.82}
V\left(a_{min}\right)  =-  \frac{1}{2}\left(\Omega_{k}  -  2\left(\left|\Omega_{m,\Lambda} \Omega_{dr} \right|\right)^{1/2}\right).
\eq
Thus, if $\Omega_{k} < 0$, we have $V\left(a_{min}\right) > 0$, and the potential is strictly positive, as shown by Curve (a)
in  Fig.\ref{fig18}. So, the motion now is forbidden. If $\Omega_{k} >  0$, we have $V\left(a_{mim}, \Omega^{c}_{m,\Lambda}\right) = 0$, 
where $\left|\Omega^{c}_{m,\Lambda}\right| = \Omega^{2}_{k}/(4\left|\Omega_{dr}\right|$, as shown by Curve (c) in Fig.\ref{fig18}. Then,
the only possible motion is that the universe is static and stays at the point $a = a_{min}$, at which we have $\dot{a} = \ddot{a} = 0$. So,
 it represents a stable point. When  $\left|\Omega_{m,\Lambda}\right| > \left|\Omega^{c}_{m,\Lambda}\right|$, the potential is always 
 positive, as shown by   Curve (b) in Fig.\ref{fig18}, so the motion in this subs-case is forbidden.  When  $\left|\Omega_{m,\Lambda}\right| <
 \left|\Omega^{c}_{m,\Lambda}\right|$, the potential is given by  Curve (d) in Fig.\ref{fig18}, from which we can see that the motion is
 restricted to the region $a \in [a_{1}, a_{2}]$, and the universe is oscillating between this two turning points, and no spacetime singularities
 is formed during the whole process.

\subsubsection{$w < -1$}

 When $w <  -1$,  Eqs.(\ref{3.56a}) and (\ref{3.56b}) can be written as 
\bqn
\lb{3.83a}
V(a) & =& - \frac{1}{2}\left(\Omega_{k}  + \frac{\Omega_{dr}}{a^{2}}\right.\nb\\
& &\left.  - a^{2}\left(\left|\Omega_{\Lambda}\right| - a^{3(|w| -1)} \Omega_{m}\right)\right), \\
\lb{3.83b}
V'(a) & =&  a\left(\left|\Omega_{\Lambda}\right|  + \frac{\Omega_{dr}}{a^{4}} \right.\nb\\
& & \left. -  \frac{(3|w|-1)\Omega_{m}}{2}a^{3(|w| - 1)}\right). 
\eqn
 
{\bf Case C.7.1) $ \; \Omega_{dr} > 0$}: In this case we have 
\bq
\lb{3.84}
V(a) = \cases{-  \infty, & $a = 0$,\cr
- \infty, & $a = \infty$.\cr}
\eq
It can also be shown that the equations $V(a) = 0 = V'(a)$ have a unique solution $(a_{max}, \Omega^{c}_{\Lambda})$. The potential
in the cases,   $\left|\Omega_{\Lambda}\right| > \left|\Omega^{c}_{\Lambda}\right|$, 
 $\left|\Omega_{\Lambda}\right| = \left|\Omega^{c}_{\Lambda}\right|$,
and  $\left|\Omega_{\Lambda}\right| < \left|\Omega^{c}_{\Lambda}\right|$, is given, respectively, by  Curves (a), (b), and
(c) in Fig. \ref{fig11}, and the corresponding motion of the universe can be easily deduced from there, as described by Fig.\ref{fig13}.

{\bf Case C.7.2) $ \; \Omega_{dr} < 0$}: In this case we have 
\bq
\lb{3.85}
V(a) = \cases{ \infty, & $a = 0$,\cr
- \infty, & $a = \infty$,\cr}
\eq
and the equations $V(a) = 0 = V'(a)$ have a unique solution $(a_{max}, \Omega^{c}_{\Lambda})$. The potential
in the cases,   $\left|\Omega_{\Lambda}\right| > \left|\Omega^{c}_{\Lambda}\right|$,  $\left|\Omega_{\Lambda}\right| 
= \left|\Omega^{c}_{\Lambda}\right|$,
and  $\left|\Omega_{\Lambda}\right| < \left|\Omega^{c}_{\Lambda}\right|$, is given, respectively, by  Curves (a), (b), and
(c) in Fig. \ref{fig16}, and the corresponding motion of the universe can be easily deduced from there.

\section{Conclusions}
\renewcommand{\theequation}{6.\arabic{equation}} \setcounter{equation}{0}

In this paper, we have studied the thermodynamics of cosmological models in  
 the Horava-Lifshitz theory of gravity, and found that the first law of thermodynamics
holds only for the perfect fluid that satisfies the condition (\ref{4.15}).

In the Horava-Lifshitz theory of gravity, the Friedmann-like equations are given by Eqs.(\ref{3.3a}) 
and (\ref{3.3b}). We have studied systematically
 these equations coupled with a perfect fluid with the equation of state $p = w\rho$, where
$p$ and $\rho$ are the pressure and energy density of the fluid, and $w$ is a constant. In this case, the 
corresponding cosmological models contain four free parameters, $\Omega_{m}, \; \Omega_{k}, \; \Omega_{dr}$ and 
$\Omega_{\Lambda}$, where both $ \Omega_{k}$ and  $\Omega_{dr}$ are proportional to the curvature
of the three-dimensional space [cf. Eq.(\ref{3.8})], and are all zero when the curvature vanishes. Then, 
the models reduces to
those given in Einstein's theory of gravity, which has been systematically studied and classified recently in \cite{Ha09}.
Therefore, in this paper we have assumed that $ \Omega_{k} \Omega_{dr} \not= 0$, but kept their signs
arbitrary. The term $\Omega_{\Lambda}$ is related to the cosmological constant, and we have studied
all the three possibilities, $\Omega_{\Lambda} =0$, $\Omega_{\Lambda} > 0$, and $\Omega_{\Lambda} <0$,
although we have assumed that $\Omega_{m} > 0$, as the latter represents the matter. 
Then,  depending on particular values of the four free parameters, we have divided the 
models  into various cases. In each case  the main properties of the evolution have been studied in detail, including 
the periods of deceleration and/or acceleration, and the existence of big bang, big crunch, and big rip singularities.  

As first noticed in \cite{brand},  models that represent a bouncing universe can be constructed, due to the presence
of the dark radiation term. However, it should be noted that the condition $\Omega_{dr} < 0$ for the existence of such models
is only a necessary condition, but not sufficient. For example, in Case c) where $\Omega_{\Lambda} < 0$, no such
models exist for all the sub-cases with $w > -1$. In addition, in order to have a bouncing universe, it is also important
that $w$ has to be less than $1/3$, so no source of matter will redshift faster than that of the dark radiation. 
However, as argued in \cite{MNTY}, the radiation in the UV regime scales as $a^{-6}$. Thus, in order to obtain 
a bouncing universe, one might need to consider the more general case studied in \cite{SVW}, in which a term
that proportional to $a^{-6}$ exists. So, if this term dominates the radiation, a bouncing universe still exists.

It should be also noted that  many of these models may not be consistent  with current observational data sets \cite{Obs}.
As a matter of fact, if we compare them with the $\Lambda$CDM model, we find that 
\bqn
\lb{6.1}
V_{\Lambda CDM}(a) &=& - \frac{1}{2} \left(\Omega_{k} + \Omega_{\Lambda}a^{2} + \frac{\Omega_{m}}{a}\right)\nb\\
&=& \cases{-\infty, & $a = 0$,\cr
-\infty, & $a = \infty$, \cr}
\eqn
which is schematically given by one of the curves given in Fig.\ref{fig9} or Fig.\ref{fig11} except for the case of Curve
(e) given there. From these curves we can see that there exists a maximal point $a= a_{max}$, for which when 
$a < a_{max} $ the universe is decelerating, and when $a > a_{max} $ it is accelerating. From Eq.(\ref{6.1})
we find that
\bq
\lb{6.2}
{V'}_{\Lambda CDM}(a)= - \frac{ \Omega_{\Lambda}}{a^{2}} \left(a^{3} - \frac{\Omega_{m}}{2 \Omega_{\Lambda}}\right).
\eq
 Since our universe now is accelerating, and recall that we have set the current radius $a_{0}$ of our   universe to one,
 from the above expression we can see that the current acceleration of the universe requires that
 $\Omega_{m} <  2 \Omega_{\Lambda}$. In fact, numerical fitting of the model shows that $\Omega_{m} \simeq 0.27$
 and $\Omega_{\Lambda} \simeq 0.73$ \cite{Obs}, for which we have $a_{max} \simeq  0.57$. 
However, we do  hope that such a classification is useful for the future studies of cosmology in the Horava-Lifshitz 
theory, as such studies  are  just starting \cite{Cosmos}.   

Finally, we note that in brane world scenarios \cite{branes}, dark radiation term also appears. In the later times of the 
universe, the quadratic terms of the energy density can be neglected, and the corresponding Friedmann equation will 
reduce to the ones studied in this paper. So, our classification presented here is also applicable to these models, too.

\begin{acknowledgements}
The authors thank R. Maartens for valuable discussions and suggestions. This work was partially supported by the NSFC
 grant, No. 10703005 and No. 10775119.
\end{acknowledgements}


\end{document}